\documentclass[lettersize,journal]{IEEEtran}

\makeatletter
\def\ps@IEEEtitlepagestyle{
  \def\@oddfoot{\mycopyrightnotice}
  \def\@evenfoot{}
}
\def\mycopyrightnotice{
  {\footnotesize
  \begin{minipage}{\textwidth}
  \centering
© 20XX IEEE.  Personal use of this material is permitted.  Permission from IEEE must be obtained for all other uses, in any current or future media, including reprinting/republishing this material for advertising or promotional purposes, creating new collective works, for resale or redistribution to servers or lists, or reuse of any copyrighted component of this work in other works.
  \end{minipage}
  }
}

\usepackage{blindtext, graphicx}
\usepackage{subcaption}
\usepackage{amsmath}
\usepackage{amsthm}
\usepackage{mwe}
\usepackage{tikz}

\usetikzlibrary{shapes,arrows}
\usepackage{array, tabularx}
\usepackage{multirow,multicol, makecell, booktabs}

\usepackage{pifont}
\usepackage{array,tabularx}
\usepackage{lipsum}
 
\usepackage{xcolor}
\usepackage{caption}
\usepackage{subcaption}
\usepackage{authblk}
\usepackage{bbm}
\usepackage{etoolbox}%
\usepackage[utf8]{inputenc}
\usepackage[english]{babel}

\usepackage[colorinlistoftodos]{todonotes}
\newcolumntype{P}[1]{>{\centering\arraybackslash}p{#1}}
\newcolumntype{C}[1]{>{\centering\arraybackslash}m{#1}}
\usepackage{amstext}  
\usepackage{footnote}
\makesavenoteenv{tabular}
\makesavenoteenv{table}

\usepackage{times}
\usepackage{fancyhdr,graphicx,amssymb}
\usepackage[linesnumbered,ruled,vlined]{algorithm2e}
\SetKw{KwBy}{by}
\usepackage{tikz}

\usepackage{float}
\floatstyle{plaintop}
\restylefloat{table}
\graphicspath{ {./images/} }
\graphicspath{{Pictures/}} 
\usepackage{tikz}

\SetCommentSty{mycommfont}
\include{pythonlisting}

\begin{document}
\title{Hybrid Non-Binary Repeated Polar Codes}


\author{\IEEEauthorblockN{Fariba Abbasi, \textit{Member, IEEE}, Hessam Mahdavifar, \textit{Member, IEEE}, and Emanuele Viterbo, \textit{Fellow, IEEE}
\thanks{This paper was presented in part at \em{IEEE International Symposium on Information Theory (ISIT)}, Melbourne, Australia, July, 2021, \cite{FaribaISIT}. \\
  Fariba Abbasi was with the Department of Electrical and Computer Systems Engineering (ECSE), Monash University, Melbourne, VIC3800, Australia. She is now with the Department of Electrical Engineering and Computer Science (EECS), University of Michigan, Ann Arbor, MI 48104 (E-mail: fabbasia@umich.edu). \\
Emanuele Viterbo is with the Department of Electrical and Computer Systems Engineering (ECSE), Monash University, Melbourne, VIC3800, Australia (E-mail: emanuele.viterbo@monash.edu).\\
 Hessam Mahdavifar is with the Department of Electrical Engineering and Computer Science (EECS), University of Michigan, Ann Arbor, MI 48104 (E-mail: hessam@umich.edu).}\\} }

\maketitle

\begin{abstract}
Concatenating the state-of-the-art codes at moderate rates with repetition codes has emerged as a practical solution deployed in various standards for ultra-low-power devices such as in Internet-of-Things (IoT) networks. In this paper, we propose a novel concatenation mechanism for such applications which need to operate at very low signal-to-noise ratio (SNR) regime. In the proposed scheme, the outer code is a hybrid polar code constructed in two stages, one with a binary kernel and another also with a binary kernel but applied over a binary extension field.  The inner  code  is  a  non-binary multiplicative  repetition  code. This particular structure inherits low-complexity decoding structures of polar codes while enabling   concatenation with an inner non-binary multiplicative repetition scheme. The decoding for the proposed scheme is done using cyclic redundancy check (CRC) aided successive cancellation list (SCL) decoder over additive white Gaussian noise (AWGN) and Rayleigh fading channels. 
Simulation results demonstrate that the proposed hybrid non-binary repeated polar code provides performance gain compared to a  polar-repetition scheme with comparable decoding complexity.

\end{abstract}

\section{Introduction}
\label{introduction}

The Third Generation Partnership Project (3GPP) has recently introduced Narrow-Band Internet-of-Things (NB-IoT) and enhanced Machine-Type Communications (eMTC) features into the cellular standard protocols. These two narrow-band and complementary technologies expand the cellular networks to support low-power, wide-area (LPWA) cellular connectivity for a wide range of IoT use cases \cite{RRatasuk}. 

In general, IoT devices need to operate under extreme power constraints. Consequently, they often communicate at very low signal-to-noise ratio (SNR), e.g., $-13$ dB or $0.03$ bits per transmission (translated to capacity) in NB-IoT protocols \cite{RRatasuk}. Also, they are often not equipped with advanced transceivers due to cost constraints. Therefore, the solution adopted in the standard is to use the legacy turbo codes or convolutional codes at moderate rates, e.g., the turbo code of rate $1/3$, together with many repetitions, e.g., up to $2048$ repetitions in NB-IoT. This implies effective code rates as low as $1.6 \times 10^{-4}$ are supported in such protocols. This repetition scheme has efficient implementations with computational complexity and latency effectively reduced to that of the outer code. However, it is expected that repeating a high-rate code to enable low-rate communication will result in rate loss and mediocre performance. As a result, studying  ultra-low-rate error-correcting codes for reliable communications in such low-capacity regimes becomes necessary \cite{MFereydounian}.


In \cite{MFereydounian}, the authors constructed an efficient repetition scheme with outer polar codes and showed that the proposed polar-repetition scheme outperforms the Turbo-repetition code, the proposed code design in the eMTC and NB-IoT (uplink) standards, over additive white Gaussian noise (AWGN) channel. In another related work, low-rate codes for binary symmetric channels were constructed by concatenating high-rate polar codes with repetitions \cite{DumerBSC}. Also, non-binary LDPC codes concatenated with multiplicative repetition codes were introduced in \cite{Kasai}. By multiplicatively repeating the $(2,3)$-regular non-binary LDPC mother code of rate $1/3$, they constructed rate-compatible codes of lower rates $1/6, 1/9, 1/12, \ldots$ which outperform the best low-rate binary LDPC codes at the cost of the increase in decoding complexity. Very recently, weakly-coded binary LDPC type code combined with polar code has been introduced in \cite{Dumerldpc}, \cite{Dumer} which is shown to outperform uncoded modulation over high noise memoryless channels. However, the complexity of the proposed scheme in \cite{Dumerldpc} and \cite{Dumer} is higher than that of a repetition-based scheme.

In \cite{Fariba}, we proposed {\em polar coded repetition}  to improve the asymptotic achievable rate of the repetition scheme, while ensuring that the overall encoding and decoding complexity is kept almost the same. A slightly modified polar codeword is transmitted in each repetition block by deviating from Ar{\i}kan's standard $2 \times 2$ kernel in a certain number of polarization recursions at each repetition block. It is shown that the polar coded repetition outperforms the  polar-repetition, in terms of the asymptotic achievable rate, for any given number of repetitions over the binary erasure channel (BEC).

In this paper, we propose an alternative mechanism for polar-repetition schemes, referred to as hybrid non-binary multiplicative repetition. In this scheme, the outer code is a hybrid binary and non-binary polar code constructed in two stages. The first stage of the outer encoder utilizes Ar{\i}kan's binary polarization kernel applied recursively, as in original polar codes \cite{Arikan}. The output bits of the first stage are grouped into $t$-tuples and are turned into symbols over the extension binary field $GF(2^t)$. Then Ar{\i}kan's kernel is again applied recursively. Hence, the output of the outer encoder consists of symbols over $GF(2^t)$. The inner code is a non-binary multiplicative repetition code.

The motivation behind constructing the outer code in two stages is to improve the decoding performance by capturing the correlation of the bits within each symbol in the decoding process.  The encoded symbols can be either turned into binary strings for transmission using a binary modulation scheme, e.g.,  binary  phase  shift  keying  (BPSK) or, alternatively, can be sent using a higher order modulation. The proposed structure allows for i) benefiting from the multiplicative repetition over an extension field as opposed to a simple repetition in a  binary polar-repetition scheme, and ii) keeping the complexity of the encoder/decoder almost the same as that of the polar-repetition scheme. 
The simulation results show that with increasing list size in the CRC-aided SCL decoder, the proposed hybrid non-binary repeated polar code provides significant performance gain compared to a  polar-repetition scheme with comparable decoding complexity over AWGN and Rayleigh fading channels. Alternatively, the proposed scheme results in lower decoding complexity compared to the polar-repetition scheme with the same performance.

The rest of this paper is organized as follows. In Section II, we review some basics of polar codes as well as repetition codes. In Section III, we discuss the proposed scheme. The numerical results are discussed in Section IV. Finally, Section V concludes the paper.

\section{Background}

In this section,
a brief background on polar codes, their {\em successive cancellation} (SC) and {\em cyclic redundancy check-aided successive cancellation list} (CRC-aided SCL) decoders, as well as some background on repetition schemes, a simple and efficient scheme for low-capacity applications, are discussed.


\subsection{Polar Codes}

An $(n, k, \mathcal{F})$, with $n=2^m$, polar code based on the $2 \times 2$ polarization kernel $G_2=\footnotesize \begin{pmatrix}
1 & 0\\
1 & 1
\end{pmatrix}$ is a linear block code generated by $k$ rows of $G_n=G_2^{\otimes m}$, where $.^{\otimes m}$ is the $m$-th Kronecker power of a matrix \cite{Arikan}. 
The set of frozen bits $\mathcal{F}$, with $|\mathcal{F}|=n-k$, is the set of the indices of sub-channels with the lowest reliabilities. Ar{\i}kan's polar codes are constructed by setting entries of the input vector $u_0^{n-1}$\footnote{$u_0^{n-1}$ is a row vector $(u_0, u_1, \ldots u_{n-1})$ and $u_0^i$ is its subvector $(u_0, u_1, \ldots u_i)$.} indexed by $\mathcal{F}$ to 
some pre-defined values, e.g. zero, and the remaining $k$ bits are used to transmit the information. At the decoder side, the SC decoder, makes the decision on $u_i$, based on the previously decoded bits, $\hat{u}_0^{i-1}$, and channel output vector, $y_0^{n-1}$, according to the following log likelihood ratio (LLR) rule:
\begin{equation}
\hat{u}_i=
\begin{cases}
1, \  \  \   \ \  \  \  \  \ \ \  \ \ \ \ \ \ \ \ \ \ \ \ \ \ \ \ \ \ \textup{if} \ \ \ i \in \mathcal{F}^C \ \& \\  \ \ \ \ \ \ \ \ \ \ \ \ \ \ \ \ \ \ \ \ \ \ \ \ \ \  \ln \frac{W_{n}^{(i)}(\hat{u}_0^{i-1},y_0^{n-1}|u_i=0)}{W_{n}^{(i)}(\hat{u}_0^{i-1}, y_0^{n-1}|u_i=1)} < 0\\
\text{the frozen value of $u_i$},  \ \  \   \  \textup{if} \ \ \ i \in \mathcal{F},  \label{uhat}
\end{cases}
\end{equation}
where $W_{n}^{(i)}$ is the $i$-th bit-channel \cite{Arikan}.


In order to improve the error correction performance of the SC decoder, the successive cancellation list (SCL) decoding algorithm was proposed by Tal and Vardy in \cite{Tal}. In SCL decoding, the $L$ most likely paths $u_0^{i-1}$ are tracked. When decoding $u_i$, for $i \in \mathcal{F}^C$, the decoder extends each path into two paths exploring both possibilities $u_i=0$ and $u_i=1$. If the number of obtained paths exceeds $L$, the decoder picks $L$ most likely paths as the surviving ones and prunes the rest based on a certain Path Metric (PM). Let $\hat{u}_i[l]$ denote the estimate of $u_i$ in the $l$-th path, for $l \in \{1, 2, \ldots, L\}$ and $S_{i}[l]$ denote its corresponding LLR. Then the corresponding PM is calculated as follows:
\begin{equation}
PM_l^{(i)}=
\begin{cases}
PM_l^{(i-1)}+|S_{i}[l]| \   \ \textup{if}   \ \hat{u}_i[l] \neq \frac{1}{2} (1-\textup{sgn} (S_{i}[l]))\\
PM_l^{(i-1)} \ \ \ \ \ \ \ \  \ \ \  \   \  \textup{otherwise},  \label{pathmetric} 
\end{cases}
\end{equation}
where $PM_l^{(-1)}$ is set to $0$. Finally, the path with the smallest PM is selected as the estimated bits $\hat{u}_0^{n-1}$. As the list size $L$ grows large, the SCL decoder approaches the maximum-likelihood (ML) decoding performance. To further improve the performance of the SCL decoder, $p$-bits cyclic redundancy check (CRC) are appended to the information bits as an outer code. 
The SCL decoder outputs the decoding path with the smallest PM among the paths which pass the CRC \cite{CRC}. Remarkably, Polar-CRC schemes have been adopted for control channels in the fifth generation of wireless networks (5G) due to demonstrating the state-of-the-art performance for a wide range of code parameters \cite{5Gspec}. Polar codes and polarization phenomenon have been also successfully applied to a wide range of problems including data compression~\cite{Arikan2,abbe2011polarization}, broadcast channels~\cite{mondelli2015achieving,goela2015polar}, multiple access channels~\cite{STY,MELK}, joint detection and decoding~\cite{Jalali}, physical layer security~\cite{MV}--\cite{wei2016polar}, fading channels ~\cite{Si}--\cite{Trifonovfading}, and  coded modulation~\cite {Seidl,MahdavifarBIM}.

\subsection{Repetition Schemes}\label{Repetition}
In low-capacity applications, a straightforward way of designing practical low-rate codes, that is also adopted in the standard, is through repetition schemes. Let $r$ denote the number of the repetitions and $N$ denote the total length of the code. For constructing a repetition scheme, first, a smaller outer code (e.g., a polar code) of length $n=N/r$ is designed and then each of its coded bits is repeated $r$ times. 
The main advantage of repetition schemes is the relatively low computational complexity of their encoding and decoding algorithms, especially when $r$ is large. This is because the encoding/decoding complexity is effectively reduced to that of the outer code. Once the outer code is constructed, at the encoder, one just needs to repeat each of the coded bits $r$ times. At the decoder, the LLR of a coded bit repeated $r$ times is equal to the sum of the $r$ LLRs of the individual transmission. While this looks promising, one should note that the main drawback of repetition schemes is the loss in capacity. In general, we have $C(W^r) \leq rC(W)$, where $W^r$ is the channel resulting from repeating $r$ times transmission across $W$, and $C(\cdot)$ is the channel capacity. Note also that the ratio $C(W^r) / rC(W)$ vanishes as $r$ grows large. Consequently, for very large $r$, the repetition code might suffer from an unacceptable rate loss \cite{MFereydounian}.

\section{The Proposed Scheme}

As mentioned in Section II-B, the polar-repetition code is a practical scheme for low-capacity scenarios due to its low encoding/decoding complexity. However, from a channel coding prospective, repeating a high rate polar code to enable low-rate communication is sub-optimal \cite{MFereydounian}. To improve the performance of the polar-repetition scheme in low-capacity regime, we propose the {\em hybrid non-binary repeated polar code} scheme: the concatenation of a binary polar code with a non-binary random multiplicative repetition code.
In this scheme, the codewords are not generated by simple repetitions of the polar code but by random multiplicative repetitions of the non-binary symbols. Our simulation analysis of the proposed scheme under SC and CRC-aided SCL decoders shows that concatenating with random non-binary multiplicative repetition code improves the polarization rate of the code and reduces the number of the low-weight codewords (see Sec. IV). 

A potential solution to benefit from the random  non-binary multiplicative code would be to concatenate it with {\em non-binary} polar codes. A similar solution is proposed in the context of LDPC coding, where gains are observed with multiplicative random non-binary LDPC codes compared to binary LDPC codes in the low-rate regime [7]. However, in the context of polar coding, the decoding complexity of non-binary polar codes is often very high for practical scenarios such as IoT, where the devices are required to be low complexity and low cost. Therefore, in this paper, we consider the {\em binary} polar codes as the outer code. In general, a straightforward way of concatenating a binary polar code with a non-binary one would result in a loss of the correlation information between the bits in a symbol and render the advantage of the non-binary outer code. However, our proposed scheme overcomes this problem by splitting the outer code into two stages to capture such correlation in the decoding process.

\begin{figure*}[h]
\includegraphics[width=1\linewidth]{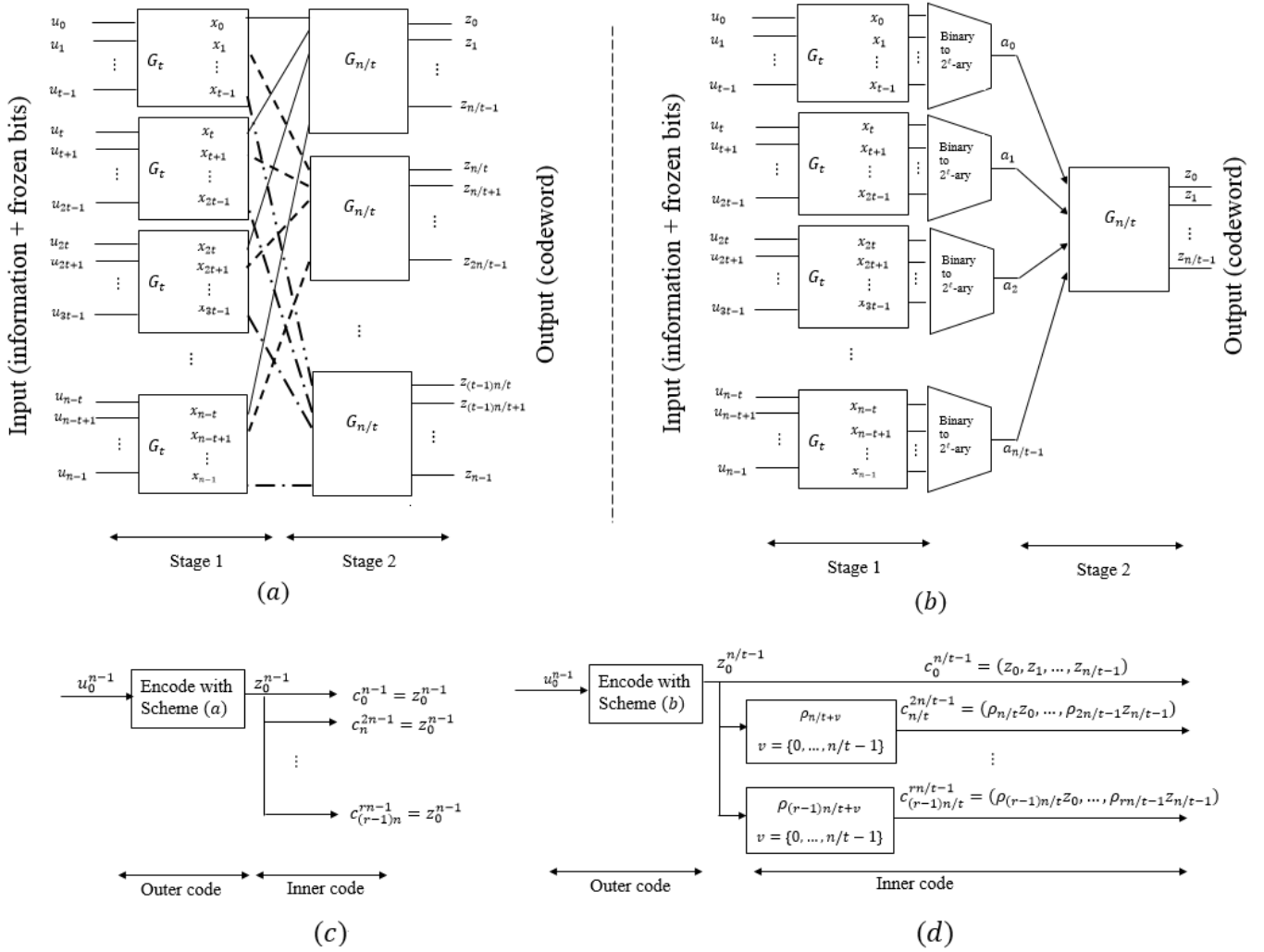}

\caption{(a) Layered factor graph representation of a polar code (b) Layered factor graph representation of the proposed hybrid non-binary polar code (c) Polar-repetition scheme (d) Hybrid non-binary repeated polar codes scheme.}
\label{fig:Concatenation}
\end{figure*}

Owing to the recursive structure of the polar codes, one can consider the polarization transform kernel $G_n$ as the concatenation of $G_{n/t}$ and $G_t$, with $G_n=G_{n/t} \otimes G_t$, where $t=2^{m^{\prime}}, m^{\prime}=\{1, 2, 3, \ldots, \}$, \cite{Mahdavifar}. Figure \ref{fig:Concatenation} (a) shows the block diagram of the encoder of the polar code with this structure, where $x_0^{n-1}$ is the output of the transformation $G_t$. Figure \ref {fig:Concatenation} (c) shows the structure of the polar-repetition. In this scheme, the outer code is the polar code depicted in Figure \ref {fig:Concatenation} (a) and the inner code is the repetition code which repeats the output of the outer polar code, $z_0^{n-1}$, $r$ times and generates code $C_r$ as follows. 
\begin{equation}
    \begin{aligned}
C_r &=\{c_0^{rn-1} | c_{(r-1)n+v}=z_v, \\
 \textup{for} & \ \ v=\{0, \ldots, n-1\}, \ \ c_0^{(r-1)n-1} \in C_{r-1}\}.
     \end{aligned}
\end{equation}
To improve the performance of the polar-repetition, the scheme depicted in Figure \ref{fig:Concatenation} (b) and (d) is proposed. The encoder and decoder of the proposed hybrid non-binary repeated polar codes are as follows.

\subsection{Encoding of the Proposed Hybrid Non-Binary Repeated Polar Codes}

For encoding the proposed scheme, in the first stage, the binary input bits $u_0^{n-1}$ are divided into subsets of bits of size $t$. Then, each of these $n/t$ $t$-tuples are encoded with binary polarization kernel $G_t$ over $GF(2)$ and the output is $x_{0}^{n-1}$. In the second stage, each of these outputs are grouped together as a $2^t$-bit symbol $a_i$, $a_i=(x_{it}, x_{it+1}, \ldots x_{(i+1)t-1}) \in GF(2^t)$, $i=\{0, 1, \ldots, n/t-1\}$. Then, the symbols $a_0^{n/t-1}$ are encoded with the binary polarization kernel $G_{n/t}$ over $GF(2^t)$ and generate code $C_1=\{z_0^{n/t-1} \in GF{(2^t)}\}$. Finally, coefficients $\rho_{n/t}^{rn/t-1}$ are chosen at random from $GF(2^t) \setminus \{0\}$ and are multiplied by $z_0^{n/t-1}$ to generate the code $C_r$, $r>1$, as follows.
\begin{equation}
    \begin{aligned}
C_r &=\{c_0^{rn/t-1} | c_{(r-1)n/t+v}=\rho_{(r-1)n/t+v}z_v, \\
 \textup{for} & \ \ v=\{0, \ldots, n/t-1\}, \ \ c_0^{(r-1)n/t-1} \in C_{r-1}\}.\label{Cr}
     \end{aligned}
\end{equation}   

Note that the coefficients $\rho_{(j-1)n/t}^{jn/t-1}$, $j=\{2, 3, \ldots, r\}$ are the random multiplication coefficients for the $j$-th repetition. Algorithm 1 shows the process of encoding the proposed scheme. The inputs to this algorithm are binary input bits $u_0^{n-1}$, $t$, $n$ and $r$. The outputs is the code $C_r$. The $\gets$ symbol denotes appending an element to a list.

\begin{algorithm}
\small
\footnotesize
\DontPrintSemicolon
\SetAlgoLined
\SetNoFillComment
\SetKwInOut{Input}{input}
\SetKwInOut{Output}{output}

\Input{$u_0^{n-1}, t, n, r$}
\Output{Code $C_r$ }
 Divide $u_0^{n-1}$ into sets of $t$ bits, $u_{it}^{(i+1)t-1}$, $i=\{0, 1, \ldots, n/t-1\}$. \\

\For {$ i = 0$ \KwTo $n/t-1$}{
     $x_{it}^{(i+1)t-1} =$ Encode each sets of bits, $u_{it}^{(i+1)t-1}$, with binary kernel $G_t$ over $GF(2)$. \\
     Group $x_{it}^{(i+1)t-1}$ together to make a $t$-bit symbol $a_i$.
}   
  $C_r= \{z_0^{n/t-1}\} \gets$ Encode $a_0^{n/t-1}$ symbols with binary kernel $G_{n/t}$ over $GF(2^t)$.\\
  
  \For {$ j = 2$ \KwTo $r$}{
   $C_r \gets$ Choose $n/t$ coefficients $\rho_{(j-1)n/t}^{jn/t-1}$ uniformly at random from $GF(2^t) \setminus \{0\}$, multiply them with $z_0^{n/t-1}$.
  }
 \KwRet $C_r$ 
\caption{Encoding Algorithm of the Proposed Hybrid Non-binary Repeated Polar Codes}\label{encoder1}
\end{algorithm}

The outer encoding only requires standard binary polar encoding followed by binary to $2^t$-ary conversion. In the decoding process, and in order to capture the correlation of the bits in a symbol, the conversion from binary to $2^t$-ary is done  between Stage 1 and Stage 2 of decoding.

Note that the authors in \cite{Presman} constructed mixed kernels over alphabets of different sizes and improved the polarization properties of the kernel $G_n$. However, in this paper, by using the structure of the kernel $G_n$, we group the binary bits into symbols without modifying the polarization kernel $G_n$.

\begin{figure*}[h]
\centering
\includegraphics[width=0.8\linewidth]{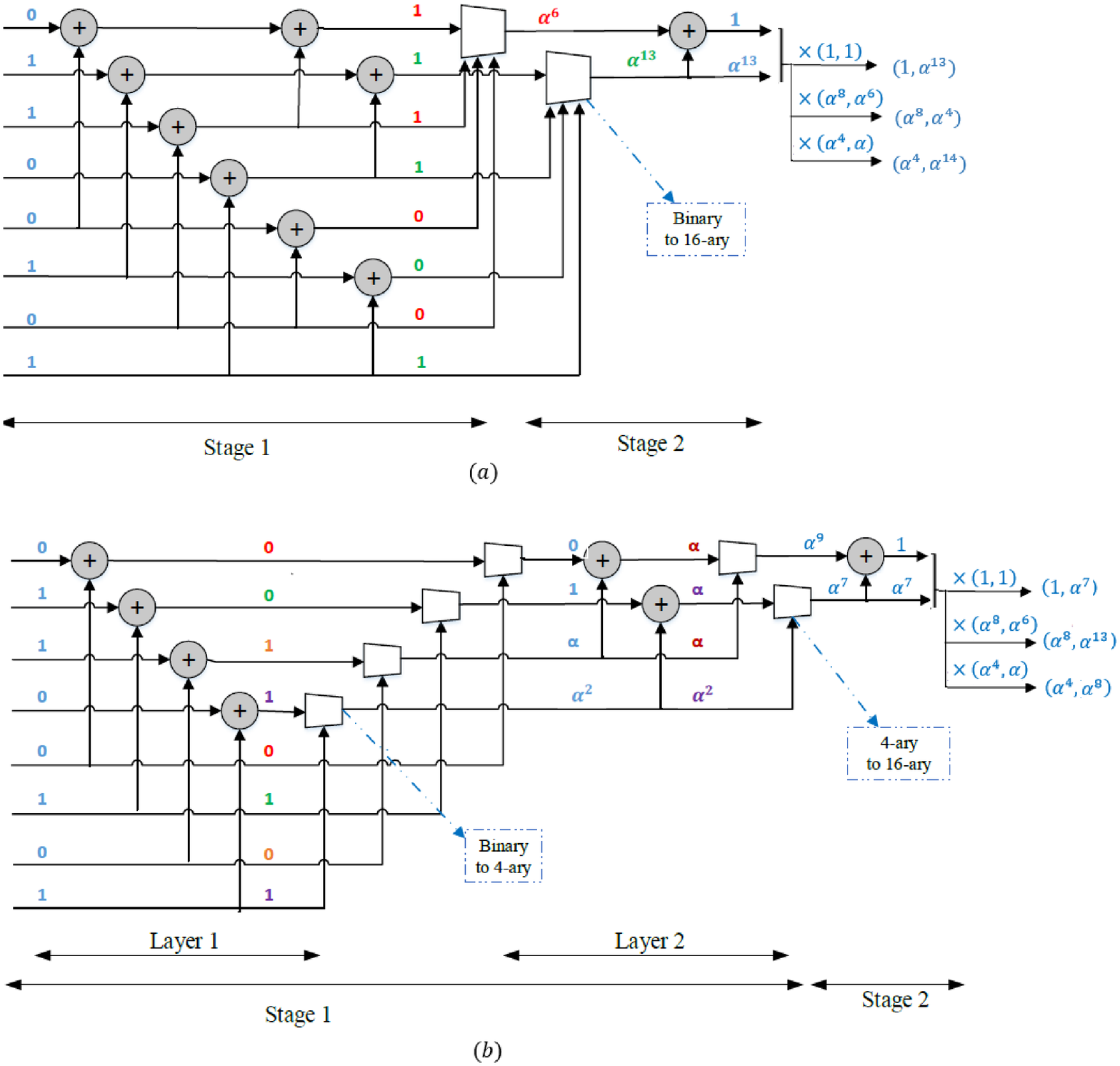}
\caption{Equivalent  encoders for the proposed scheme for $n=8$, $r=3$, $t=4$ based on (a)  Algorithm 1 and (b) Algorithm 2.}

\label{fig:example}
\end{figure*}

\textbf{Example 1:} Figure \ref{fig:example} (a) shows an example for $n=8$, $r=3$, and $t=4$. The polarization kernels of stage 1 and 2 are $G_4=B_4 G_2^{\otimes 2}$  and $G_2$, respectively, where $B_4$ is the bit-reversal permutation matrix. 
By choosing the primitive $\alpha \in GF(2^4)$ with $\pi (\alpha)=\alpha^4+\alpha+1=0$, the output codewords of this example are $c_0^{5}=(1,  \alpha^{13}, \alpha^8, \alpha^{4}, \alpha^4, \alpha^{14})$.

The encoded symbols $c_i$ are turned into binary strings for transmission over AWGN and Rayleigh fading channels using the BPSK modulation scheme, with the received signals for these two channels given by
\begin{equation*}
    y_i=(c_i)_M+w_i, \ \ \ i=\{1, 2, \ldots, N\}
\end{equation*}
and
\begin{equation*}
    y_i=h_i (c_i)_M+w_i, \ \ \ i=\{1, 2, \ldots, N\},
\end{equation*}
respectively, where $(c_i)_M$ is the modulated codeword, $h_i$ is the channel gain with Rayleigh distribution and $w_i$ is a zero mean
Gaussian noise, $w_i \sim \mathcal{N} (0, \sigma^2)$.
Here, we assume that $h_i$ and $w_i$ are independent, the receiver knows the channel state information (CSI)  and the transmitter knows only the distribution of the CSI.

In the block fading channel model, a transmission frame
of $N$ symbols is affected by $1 \leq B \leq N$ dependent fading realizations, resulting in a block of $\tau =N/B$ symbols being affected by the same fading realization. Different values $B$ represents different types of fading, e.g., $B=1$ is referred to fast fading and  $B=N$ is referred to slow fading.


\subsection{SCL Decoding of the Proposed Hybrid Non-Binary Repeated Polar Codes}
For decoding the proposed scheme, we use CRC-aided LLR-based SCL decoder. Algorithm \ref{ListDecoder} shows the details of the process. The SCL decoding of the proposed scheme involves mainly 4 parts, i.e. the initial LLRs, the LLRs of the multiplicative repetition, the LLRs of Stage $1$ and the LLRs of Stage $2$ of the outer polar code. The details of the LLRs calculations are as follow.

    \textbf{Initial LLRs:}
    The initial LLRs of the $i$-th symbol, $i \in \{0, 1, \ldots, rn/t-1\}$, with symbol value $s \in GF(2^t)$ for AWGN and block fading channels are defined as 
    \begin{equation}
        {S}_{in,i}^{(s)}=\ln{\frac{W(y_i|(c_i)_M=0_M)}{W(y_i|(c_i)_M=s_M)}},\label{initialLLRsA}
    \end{equation}
    and
     \begin{equation}
        {S}_{in,i}^{(s)}=\ln{\frac{W(y_i|(c_i)_M=0_M, h_i)}{W(y_i|(c_i)_M=s_M, h_i)}},\label{initialLLRsF}
    \end{equation}
    respectively. 
    Let us denote ${S}_{in,i}=\{S_{in,i}^{(s)}\}_{s \in GF(2^t)}$, the vector of the $i$-th initial LLRs for all possible symbols $s$ over $GF(2^t)$. 
    
   \textbf{LLRs of the multiplicative repetition:}
    Since $GF(2^t)$ is a finite field, the non-zero elements can be expressed as powers of a primitive element $\alpha$ in the field, i.e., $1,\alpha,\dots, \alpha^{2^t-2}$. Therefore, multiplication by an arbitrary non-zero symbols $\rho_j = \alpha^{\tau}$, $j=\{n/t, \ldots, rn/t-1\}$, can be regarded as a cyclic shift of the field elements by $\tau$. As a result, the decoder permutes the LLR vector ${S}_{in,i}$ and outputs the vector $\pi_{\rho_j}({S}_{in,i})$.

    Finally, the LLR of an $r$-tuple consisting of $r$ independent transmissions of symbols is equal
    to sum of the LLRs of the individual channel outcomes after the permutations (see lines 2-8 of the Algorithm \ref{ListDecoder}). 
   
    \textbf{LLRs update of Stage $2$:}
    In general, the LLR of the $i$-th symbol, $i \in \{0, 1, \ldots, n-1\}$ , with symbol value $s$ over kernel $G_n$, can be calculated according to the following formula, \cite{Arikan}:
    \begin{equation}
\begin{aligned}
 S_{i}^{(s)} & \overset{\Delta}{=} \ln \frac{\sum_{u_{i+1}^{n-1}  }R_{G_n}(\hat{u}_0^{i-1},0,u_{i+1}^{n-1})}{\sum_{u_{t+1}^{n-1}} R_{G_n}(\hat{u}_0^{i-1},s,u_{i+1}^{n-1})},\label{LLR}
 \end{aligned}
\end{equation}
where $R_{G_n}(u)=\exp(-\sum_{j=0}^{n-1}S_j^{(x_j)})$ and $x_j$ is the $j$-th index of the vector $x_0^{n-1}=u_0^{n-1}G_n$ and $s$ is an element from $GF(2^t)$. One can use the following equation for simplifying eq. (\ref{LLR}). 
\begin{equation*}
    \ln (\sum_i e^{-f_i}) \approx - \min_i (f_i).
\end{equation*}

 Now, for our binary kernel $G_2$, consider two input LLR vectors ${S}_{+}$ and ${S}_{-}$ of size $2^t$. Then,  the  output LLR vectors $\hat{S}_{+} $ and $\hat{S}_{-} $ can be derived from (\ref{LLR}), as:
 \begin{equation}
     \begin{aligned}
      \hat{S}_{+}^{(s)} \approx & \min_{u_1 \in GF(2^t)}{(S_{+}^{(s+u_1)}+S_{-}^{(u_1)})}\\
      & - \min_{u_1 \in GF(2^t)}{(S_{+}^{(u_1)}+S_{-}^{(u_1)})},\\
      \hat{S}_{-}^{(s)} \approx & ~S_{+}^{(\hat{u}_0+s)} +S_{-}^{(s)}-S_{+}^{(\hat{u}_0)}-S_{-}^{(0)}.\label{fs2}
     \end{aligned}
 \end{equation}
 where $S_{+}^{(i)}$ and $S_{-}^{(j)}$ are the $i$-th and $j$-th index of the vectors ${S}_{+}$ and ${S}_{-}$, respectively and $\hat{u}_0$ is the previously decoded symbol, while $u_1$ is the yet to be decoded symbol.

\textbf{LLRs update of Stage $1$:}
For updating the LLRs of Stage $1$, consider the input non-binary LLR vector $\hat{S}$. The output binary LLRs $\hat{S}_{b_i}$, $i=\{1, 2, \ldots, t\}$, derived from eq. (\ref{LLR}), are as follows.



\begin{equation}
\begin{aligned}
 \hat{S}_{b_i} &  \approx   \min_{u_{i+1}^{t-1}} \hat{S}^{([v^{(1)}]_{2^t})} -\min_{u_{i+1}^{t-1}} \hat{S}^{([v^{(0)}]_{2^t})},\label{LLRstage1}
 \end{aligned}
\end{equation}
where $[v^{(\beta)}]_{2^t}$ is the representation of the binary vector $v^{(\beta)}=(\hat{u}_0^{i-1},\beta,u_{i+1}^{t-1})G_t$ as element of $GF(2^t)$ for $\beta=\{0, 1\}$ and $\hat{S}^{([v_\beta]_{2^t})}$ is the $[v^{(\beta)}]_{2^t}$-th index of the vector $\hat{S}$.

\begin{figure}[h]
\centering
\includegraphics[width=1\linewidth]{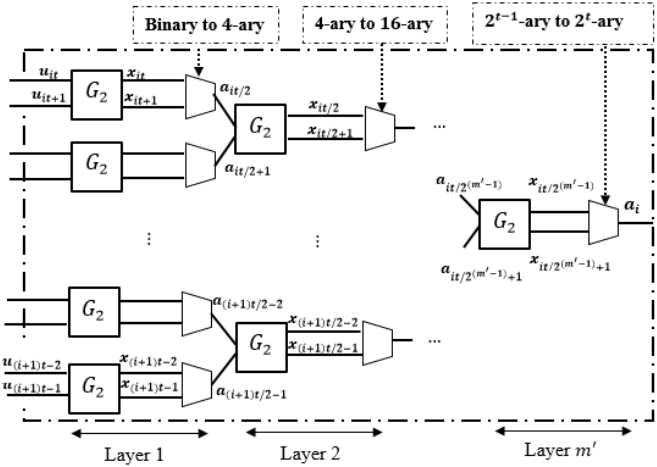}
\caption{Part of Stage 1 of the encoder based on Algorithm 2.}
\label{fig:Recurively}
\end{figure}

In Eq. (\ref{LLRstage1}) we show how to calculate the output binary LLRs $\hat{S}_{b_i}$ given the input non-binary LLR $\hat{S}$ in one step. However, updating the LLRs of Stage 1 can be done recursively as well. To this end, Stage 1 of the encoder in Figure \ref{fig:Concatenation} (b) needs to be modified. Instead of encoding $t$ bits with binary polarization kernel $G_t$ over $GF(2)$, $G_2$ is applied recursively on the inputs symbols over $GF(2^{2^{k^{\prime}-1}})$ in $k^{\prime}$ levels ($1 \leq k^{\prime} \leq m^{\prime}$). Then the obtained symbols are grouped into $2^{k^{\prime}}$-bit symbols $a_0^{(n/2^{k^{\prime}})-1}$. Finally, the output of the last level, $t$-bit symbol $a_i$, $i=\{0, 1, \ldots (n/t)-1\}$, is fed into Stage $2$. Part of Stage 1 for this modified encoder is demonstrated in Figure \ref{fig:Recurively}. The process of the encoding for this structure is presented in Algorithm \ref{encoder2}.

\begin{algorithm}
\small
\footnotesize
\DontPrintSemicolon
\SetAlgoLined
\SetNoFillComment
\SetKwInOut{Input}{input}
\SetKwInOut{Output}{output}

\Input{$u_0^{n-1}, m^{\prime}, t, n, r$}
\Output{Code $C_r$ }
 
Define List $\text{A}=\{\}$\\
$\text{A} \gets u_0^{n-1}$\tcp*{Init.}
\For {$ k^{\prime} = 1$ \KwTo $m^{\prime}$}{
\For {$ i = 0$ \KwTo $(n/2^{k^{\prime}})-1$}{
     $x_{2i}^{2(i+1)-1} =$ Encode each sets $(A[2i], A[2(i+1)-1])$ with binary kernel $G_2$ over $GF(2^{2^{k^{\prime}-1}})$. \\
     Group $x_{2i}^{2(i+1)-1}$ together to make a $2^{k^{\prime}}$-bit symbol $a_i$.
}
$A = \{\}$ \tcp*{empty List $A$}
$A \gets a_0^{(n/2^{k^{\prime}})-1}$
}
lines 6-9 of Algorithm \ref{encoder1}.\\
 \KwRet $C_r$ 
\caption{Encoding of the Proposed Hybrid Non-binary Repeated Polar Codes  for Stage $1$}\label{encoder2}
\end{algorithm}

\textbf{Example 2:} Figure \ref{fig:example} (b) demonstrates a encoder structure for the code discussed in Example 1 based on Algorithm 2. The polarization kernels of Stage 1 (Layer 1 and Layer 2) and Stage 2 is $G_2$.
The output codewords in this example are $c_0^{5}=(1,  \alpha^{7}, \alpha^8, \alpha^{13}, \alpha^4, \alpha^{8})$ over $GF(16)$.

In order to update the LLRs of the layers of Stage 1, based on Algorithm $2$, the following recursive formulas can be used for calculating $\hat{S}_{+}$ and $\hat{S}_{-}$ from $\hat{S}$.
\begin{equation}
\begin{aligned}
 \hat{S}_{+}^{(s)} &  \approx \min_{u_1 \in GF(2^{t^{\prime}})} \hat{S}^{([v^{(s)}]_{2^{t^{\prime}}})}-\min_{u_1 \in GF(2^{t^{\prime}})} \hat{S}^{([v^{(0)}]_{2^{t^{\prime}}})}\\ 
 \hat{S}_{-}^{(s)} &  \approx  \hat{S}^{([\hat{v}^{(s)}]_{2^{t^{\prime}}})}-\hat{S}^{([\hat{v}^{(0)}]_{2^{t^{\prime}}})}, \label{LLRstage1v2}
 \end{aligned}
\end{equation}
where $0 \leq  j \leq m^{\prime}-1$, $t^{\prime}=2^{j}$, $[v^{(\beta)}]_{2^{t^{\prime}}}$ is the representation of the binary vectors $v^{(\beta)}=(\beta,u_1)G_2$ as elements of $GF(2^{t^{\prime}})$, and $[\hat{v}^{(\beta)}]_{2^{t^{\prime}}}$ is the representation of the binary vectors $\hat{v}^{(\beta)}=(\hat{u}_0,\beta)G_2$ as elements of $GF(2^{t^{\prime}})$ for $\beta=\{0, s\}$.

The CRC-aided SCL decoder after updating the LLRs and calculating the PM for different paths based on the values of the $\hat{S}_{b_i}$  chooses the path with the smallest PM which passes the CRC.

\begin{algorithm}
\small
\footnotesize
\DontPrintSemicolon
\SetAlgoLined
\SetNoFillComment
\SetKwInOut{Input}{input}
\SetKwInOut{Output}{output}

\Input{List size $L$, $r$, $n$, $t$, $\rho_{n/t}^{rn/t-1}$ and $y_0^{rn/t-1}$}
\Output{The estimated bits $\hat{u}_0^{n-1}$ }
${S}_{in,i} =$ Calculate the initial LLRs with eq. (\ref{initialLLRsA}) or eq. (\ref{initialLLRsF}) for all $i=\{0, 1, \ldots, rn/t-1\}$ by using $y_0^{rn/t-1}$.\\
 
 ${S}_{\textup{inner},i} = {S}_{in,i}, i=\{0, \ldots, n/t-1\}$\tcp*{Init.: Pick the first $n/t$ elements of ${S}_{in,i}$.}
 
 \For {$ i = n/t$ \KwTo $rn/t-1$}{
 $\pi_{\rho_i}({S}_{in,i})$ \tcp*{Permute the vector ${S}_{in,i}$ based on the random coefficient $\rho_i$.}
 }
 \For{$ j = 2$ \KwTo $r$}{
 ${S}_{\textup{inner},i} = {S}_{\textup{inner},i}+\pi_{\rho_i}({S}_{in,i})[(j-1)n/t:jn/t-1]$.\tcp*{Output LLRs of the inner code}
 }

\For {$ k^{\prime} = 0$ \KwTo $n-1$ \KwBy $t$}{
     Using the LLRs ${S}_{\textup{inner},i}$, for $i=\{0, \ldots, n/t-1\}$, update the LLRs for Stage $2$ with eq. (\ref{fs2}).\\
     \For {$ j = 1$ \KwTo $t$}{
    
    Using the updated LLRs from Stage $2$, update the LLRs for Stage $1$ with eq. (\ref{LLRstage1}) and obtain $\hat{S}_{b_j}$.\\
     Calculate the PM for $\hat{S}_{b_j}$ with eq. (\ref{pathmetric}) for each of the $L$ paths.\\
     Update bits for Stage $1$.\\
    }
    Update the symbols for Stage $2$.\\
}   
 CRC-aided SCL decoder with list size $L$ chooses the path with the smallest PM which passes the CRC and outputs $\hat{u}_0^{n-1}$.\\ 
\KwRet $\hat{u}_0^{n-1}$ 

\caption{List Decoding Algorithm for the Proposed Hybrid Non-binary Repeated Polar Codes}
\label{ListDecoder}
\end{algorithm}

\section{Analysis and Numerical Results} 
In this section, we first provide complexity analysis of the proposed scheme and compare it with those of the polar-repetition scheme and the scheme in \cite{Dumerldpc}. Then, we analyze the numerical result of the proposed scheme given certain underlying binary field extensions in the construction, i.e., over GF($4$) and GF($16$), and compare it with the polar-repetition scheme and the scheme in \cite{Dumerldpc} over AWGN and Rayleigh fading channels. 
Finally, we analyze the performance of the hybrid non-binary repeated polar code.

\subsection{Complexity Analysis} \label{Complexity}

In this subsection, we count the exact number of operations needed for calculating the LLRs of the proposed scheme, polar-repetition, and the scheme in \cite{Dumerldpc} to show the performance-complexity trade-off in subsection \ref{NA}. Note that to demonstrate the computational complexity, we used the total number of the summation and comparison operations for calculating the LLRs of the outer and inner codes for the SC decoder. The complexity advantages of the entire system using CRC-aided SCL decoder will be shown in Sec. IV-B.

      \textbf{Polar-repetition code:} To update the LLRs of the  polar-repetition, one needs to compute
      \begin{itemize}
          \item $(n-1)r$ operations for decoding the inner repetition code of size $r$.
          \item $  (\frac{n}{2} \times 4+ \frac{n}{2} \times 1) \log n$ operations for decoding the outer polar code of size $n$.
      \end{itemize}
      Consequently, the total complexity is $nr+2.5 n \log n$ operations.

      \textbf{Hybrid non-binary repeated polar code:}  Since the two methods proposed in Algorithms \ref{encoder1} and \ref{encoder2} have the same computational complexities, we only count the number of the operations of the first method as follows.
    \begin{enumerate}
        \item  \textbf{Inner multiplicative repetition code:}  According to (\ref{initialLLRsA}) or (\ref{initialLLRsF}), the size of the initial LLR vector of the proposed scheme, $S_{in,i}$, is $2^t$ and $S_{in,i}^{(0)}=0$. Consequently $(2^t-1) rn/t$ operations are needed for computing  lines 6-8 of  Algorithm \ref{ListDecoder} for decoding the inner multiplicative repetition code.
        
        \item  \textbf{Stage2:} For updating the LLRs of Stage 2, one need to compute $\hat{S}_{+}^{(s)}$ and $\hat{S}_{-}^{(s)}$ given in (\ref{fs2}). 
        
       To obtain $\hat{S}_{+}^{(s)}$, one should compute
        \begin{itemize}
        \item  $S_+^{(s+u_1)}+S_-^{(u_1)}$, $\forall s, u_1 \in GF(2^t)$, with $(2^t-1) \times (2^t-1)$ operations. \footnote{Note that $S_+^{(u_1)}+S_-^{(u_1)}$ is a special case of $S_+^{(s+u_1)}+S_-^{(u_1)}$ for $s=0$ and $\hat{S}_+^{(0)}=0$, $\hat{S}_-^{(0)}=0$.}
        \item  $\min_{u_1 \in GF(2^t)}(S_+^{(s+u_1)}+S_-^{(u_1)})$, $\forall s \in GF(2^t)$, with $2^t \times (2^t-1)$.
        \item  $\hat{S}_+^{(s)}$ with $2^t-1$ operations for $s \in GF(2^t)\setminus\{0\}$.
        \end{itemize}
        Therefore, the total number of the operations for computing $\hat{S}_{+}^{(s)}$ is $(2^{t}-1)\times 2^{(t+1)}$.
        
       To obtain $\hat{S}_{-}^{(s)}$, one should compute
        \begin{itemize}
        \item $S_+^{(\hat{u}_0+s)}+S_-^{(s)}$, $\forall s \in GF(2^t)$, with $2^t-2$ operations.
        \item  $\hat{S}_{-}^{(s)}$ with $2^t-1$ operations for $s \in GF(2^t)\setminus \{0\}$. 
         \end{itemize}
        Therefore, in total $2^{t+1}-3$ operations are needed for computing $\hat{S}_{-}^{(s)}$.
       
    Finally, for the Ar\i kan's kernel of size $n/t$, 
    $(\log \frac{n}{t}) [\frac{n}{2t} ((2^{t}-1)\times 2^{(t+1)})+\frac{n}{2t} (2^{t+1}-3)]$ operations are needed for updating the LLRs of  Stage $2$.
        
        \item \textbf{Stage 1:} According to (\ref{LLRstage1}), for each $b_i, i \in \{1, 2, \ldots, t\}$, one needs to compute
        \begin{itemize}
            \item $\min_{u_{i+1}^{t-1}} \hat{S}^{([v^{(\beta)}]_{2^t})}$, $\beta=\{0,1\}$ with $2^{t-i}-1$ operations.
            \item $\hat{S}_{b_i}$ with $1$ operation.
        \end{itemize}
        As a result, in total, $\frac{n}{t} \sum_{i=1}^t (2(2^{t-i}-1)+1)$ operations are needed to update the LLRs of the ${n}/{t}$ kernel $G_t$ of size $t$. 
    \end{enumerate}
\textbf{Scheme proposed in \cite{Dumerldpc}:} This scheme is the concatenation of the $(n_l,k_l)=(8192,128)$ LDPC with $(n_p,k_p)=(128,80)$ polar code. In this scheme, one needs $8192(8\log(128)+32)=720896$ operations for decoding the LDPC code and $110469$ operations for decoding the polar code under SCL with list size $L=32$. As a result the total number of operations for this scheme is $831365$.

In a nutshell, we conclude that the complexity of the total decoding process of the polar-repetition and the proposed schemes are $O(nr+ n \log n)$ and $O(2^t rn/t+2^{2t} (n/t) \log (n/t) + 2^t n/t)$, respectively. Here, if $t$ and $r$ be  constants, the complexity of both schemes will be $O(n \log n)$. Moreover, the complexity of the scheme proposed in \cite{Dumerldpc} is $O(n_l (\log n_p +L)+Ln_p \log n_p)$. However, for numerical comparison purposes, we use the aforementioned exact number of operations for the different schemes. The complexity comparison of the three discussed schemes is shown in Table \ref{complexitycomparison} for $N=8192$ and three different values $r=16, 32, 64$.

\begin{table*}
\begin{center}
\scalebox{0.9}{
\begin{tabular}{|P{4cm}||P{3.5cm}|P{4.5cm}|P{4.5cm}|P{1cm}|}
 \hline
  \multirow{2}{*}{Parameters} & \multicolumn{4}{|P{11cm}|}{{Repetition Scheme}} \\
  \cline{2-5}
  & \multicolumn{1}{|P{2.5cm}|}{{Inner rep. $n(r-1)$ }} & \multicolumn{2}{|P{7cm}|}{{Outer polar code \ \ \ \ \ \ \ \ \  \ \ \ \ \ \ \ \ \ \ \ \ \ \ \ \ \ \ \ \ \ \ \ \ \ \ \ \ \ \ \ \ \ \ \  \ \ \ $2.5 \times n \log n$}} & \multicolumn{1}{|P{1cm}|}{{Total }}\\
\hline
 $n=512$, $r=16$ & $7680$ & \multicolumn{2}{|P{7cm}|}{$11520$} & $19200$  \\
\hline
$n=256$, $r=32$  & $7936$ & \multicolumn{2}{|P{7cm}|}{$5120$} &   $13056$     \\
\hline 
$n=128$, $r=64$ & $8064$ & \multicolumn{2}{|P{7cm}|}{$2240$} &  $10304$      \\
\hline
\hline
  \multirow{2}{*}{Parameters} & \multicolumn{4}{|P{12cm}|}{{Proposed Scheme over GF(4)}} \\
  \cline{2-5}
  & \multicolumn{1}{|P{3cm}|}{{Inner rep. 
  $\frac{n(r-1)}{t}(2^t-1)$}} & \multicolumn{1}{|P{3.5cm}|}{{Stage 2   
  $(2^{2t}-3/2)\frac{n}{t}\log \frac{n}{t}$}} & \multicolumn{1}{|P{3cm}|}{{Stage 1 \ \ \ \ \ \ \ \ \ \ \ \ \ \ \ \ \ \ \ \ \ \ \ \ \ \ $2n$}} & \multicolumn{1}{|P{1cm}|}{{Total }}\\
\hline
 $n=512$, $r=16$, $t=2$ & $11520$ & \multicolumn{1}{|P{3.5cm}|}{
 $29696$} & \multicolumn{1}{|P{3cm}|}{$1024$} & $42240$
 \\
\hline
$n=256$, $r=32$, $t=2$  & $11904$ & \multicolumn{1}{|P{3.5cm}|}{
$12992$} & \multicolumn{1}{|P{3cm}|}{$512$} & 
$25408$\\
\hline 
$n=128$, $r=64$, $t=2$ & $12096$ & \multicolumn{1}{|P{3.5cm}|}{
$5568$} & \multicolumn{1}{|P{3cm}|}{$256$}& 
$17920$\\
\hline
\hline
  \multirow{2}{*}{Parameters} & \multicolumn{4}{|P{12cm}|}{{Proposed Scheme over GF(16)}} \\
  \cline{2-5}
  & \multicolumn{1}{|P{3cm}|}{{Inner rep. 
  $\frac{n(r-1)}{t}(2^t-1)$}} & \multicolumn{1}{|P{3.5cm}|}{{Stage 2   
  $(2^{2t}-3/2)\frac{n}{t}\log \frac{n}{t}$}} & \multicolumn{1}{|P{4cm}|}{{Stage 1 $\frac{n}{t} \sum_{i=1}^{t} (2(2^{t-i}-1)+1)$}} & \multicolumn{1}{|P{1cm}|}{{Total }}\\
\hline
 $n=512$, $r=16$, $t=4$ & $28800$ & \multicolumn{1}{|P{3.5cm}|}{
 $228032$}
 & \multicolumn{1}{|P{3.5cm}|}{$3328$} & 
 $260160$\\
\hline
$n=256$, $r=32$, $t=4$  & $29760$ & \multicolumn{1}{|P{3.5cm}|}{
$97728$} & \multicolumn{1}{|P{3.5cm}|}{$1664$} & 
$129152 $    \\
\hline 
$n=128$, $r=64$, $t=4$ & $30240$ & \multicolumn{1}{|P{3.5cm}|}{
$40720$} & \multicolumn{1}{|P{3.5cm}|}{$832$}& 
$71792 $\\
\hline
\end{tabular}%
}
\end{center}
\caption{ Complexity comparison of different schemes}
\label{complexitycomparison}
\end{table*}

\subsection{Numerical Analysis}\label{NA}

In this subsection, we provide numerical results for the proposed scheme given $GF(4)$ and $GF(16)$ as the underlying binary field extension in the construction\footnote{For the simulation, Algorithm 1 is used. Although both Algorithm 1 and 2 show the same performance.}. The communication is assumed over AWGN and Rayleigh fading channels with BPSK modulation, where encoded symbols are turned into binary strings and then modulated. Also, the decoding is performed under the SC and CRC-aided SCL decoding algorithms. Then, we compare the performance of the proposed scheme with those of the polar-repetition scheme as well as the scheme proposed in \cite{Dumerldpc}. Note that the SCL decoder is implemented using the randomized order statistic algorithm for the selection of the $L$ most likely paths in each stage, which has the complexity $O(L)$, where $L$ is the list size. Moreover, the construction for the proposed hybrid non-binary repeated polar code and polar-repetition schemes is based on the Monte-Carlo simulation at an optimized design SNR, denoted by $\gamma$, to be specified separately for each case.

It is worth noting that in each instance of the simulation, a different set of the random coefficients $\rho_{(j-1)n/t}^{jn/t-1}$, $j=\{2, 3, \ldots, r\}$  is used. At the end, the average is taken over all the instances. An alternative approach is to choose the set of the coefficients in a certain way and then use them in all the instances of the simulation. The choice of such coefficients could be optimized, however, solving such a combinatorial optimization problem is complex and beyond the scope of this paper.

\begin{figure*}[tp!]
 \centering
\begin{subfigure}[t]{0.45\textwidth}
 \centering
\includegraphics[width=1.1\linewidth]{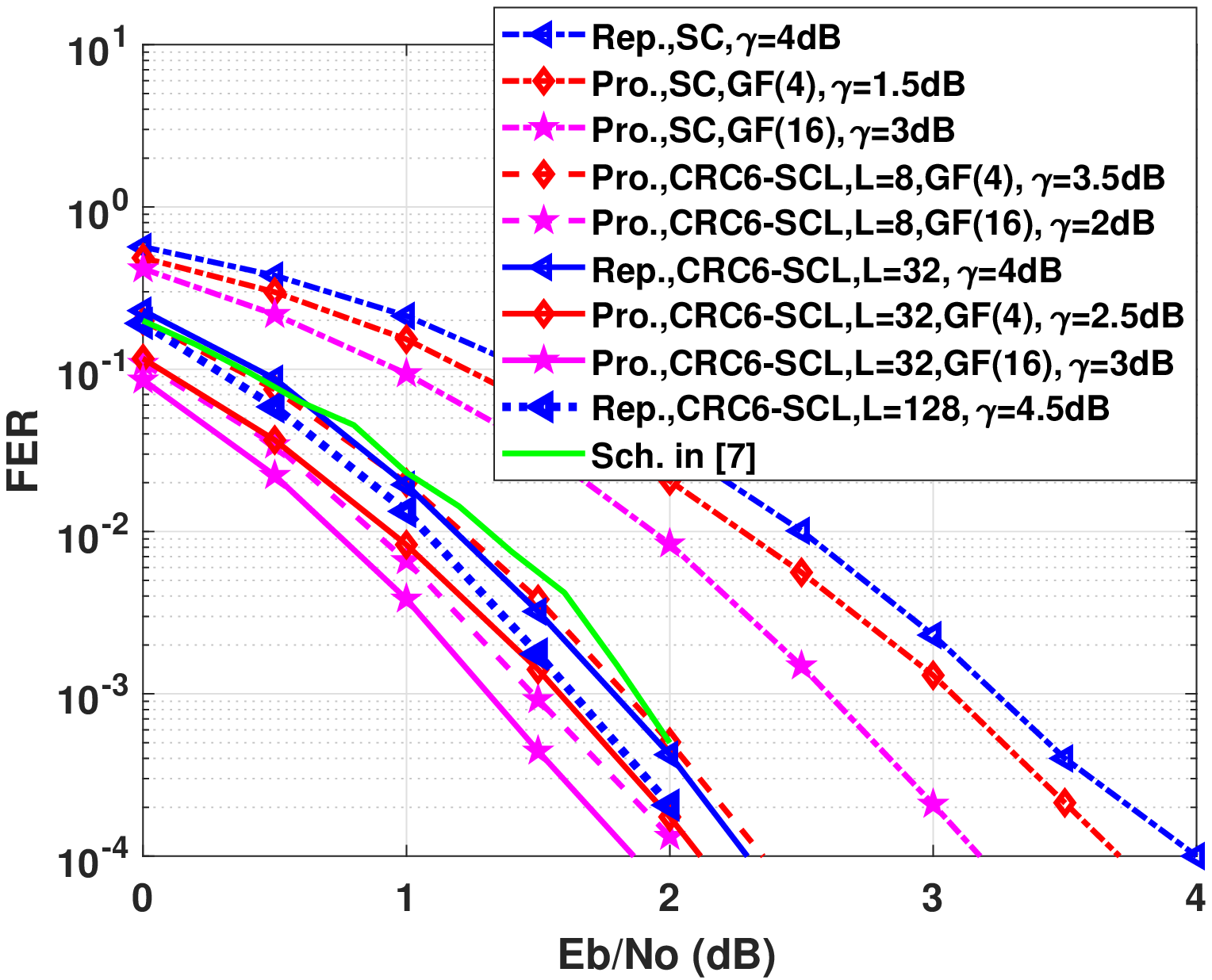}
\caption{$N=8192$, $n=512$, $k=80$, $r=16$, $t=2, 4$.}
\label{fig:512k80}
\end{subfigure}%
\begin{subfigure}[t]{0.45\textwidth}
\includegraphics[width=1.1\linewidth]{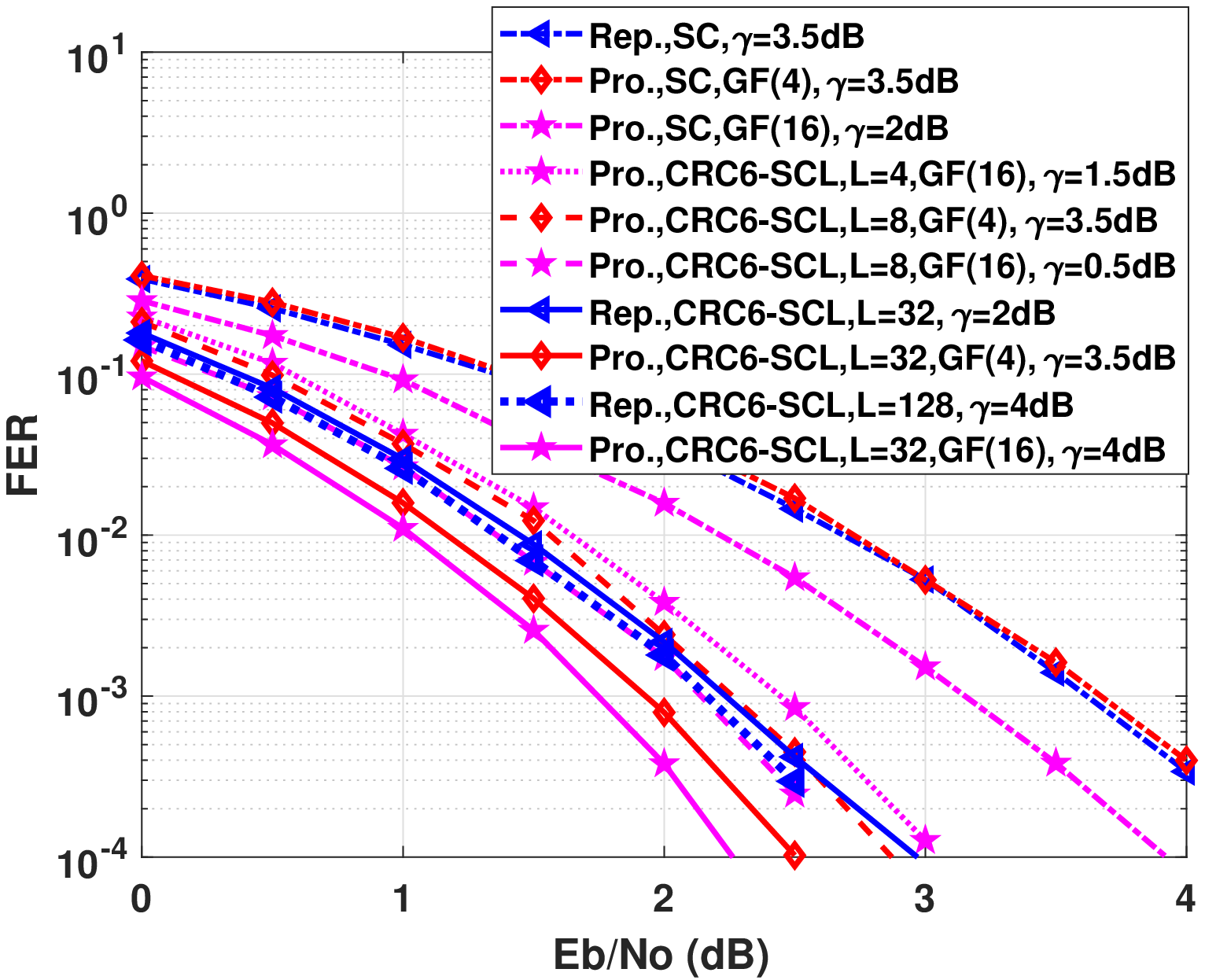}
\caption{$N=8192$, $n=512$, $k=40$, $r=16$, $t=2, 4$.}
\label{fig:512k40}
\end{subfigure}

\medskip
\begin{subfigure}[t]{0.45\textwidth}
\centering
\includegraphics[width=1.1\linewidth]{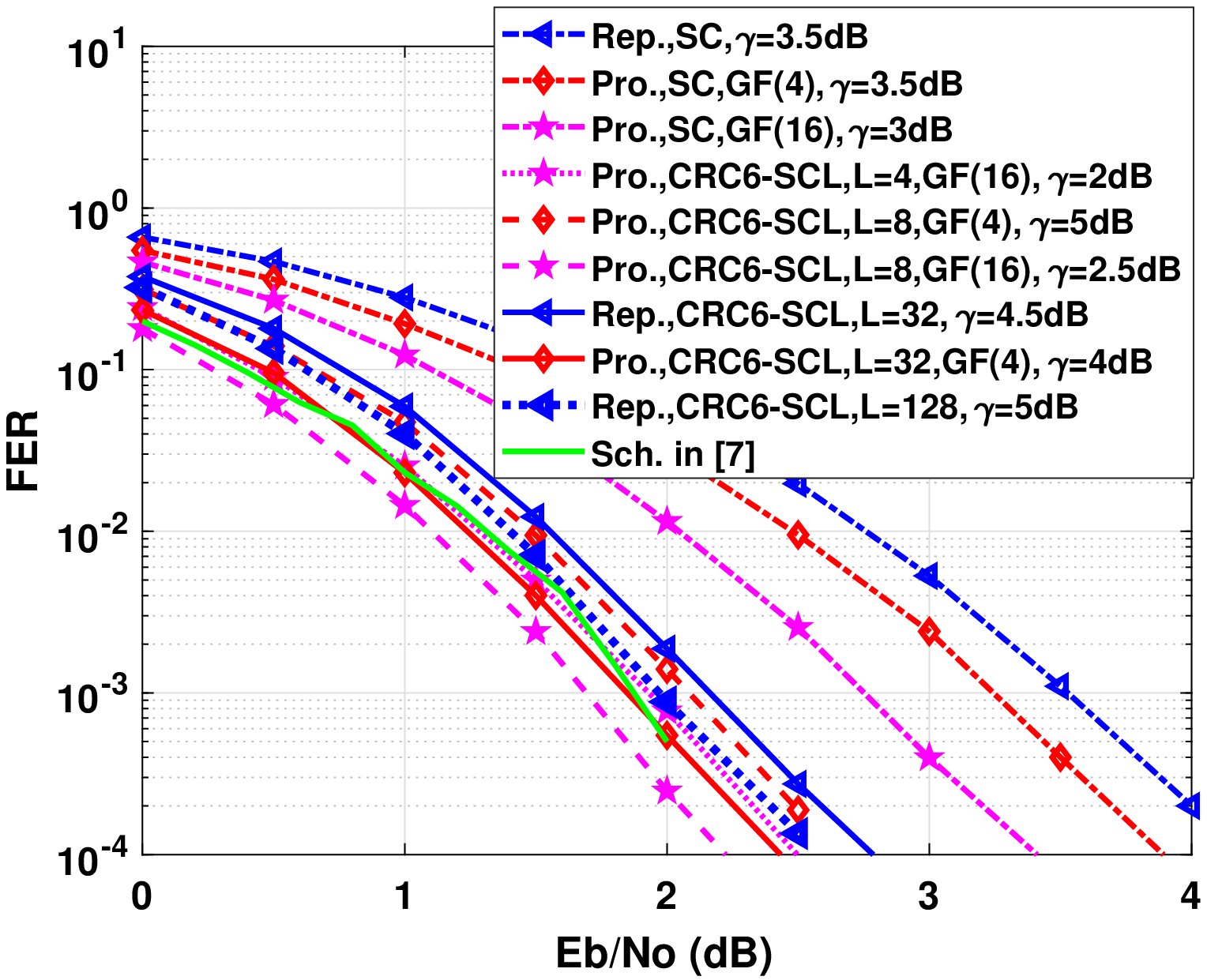}
\caption{$N=8192$, $n=256$, $k=80$, $r=32$, $t=2, 4$.}
\label{fig:256k80}
\end{subfigure}%
\begin{subfigure}[t]{0.45\textwidth}
\centering
\includegraphics[width=1.1\linewidth]{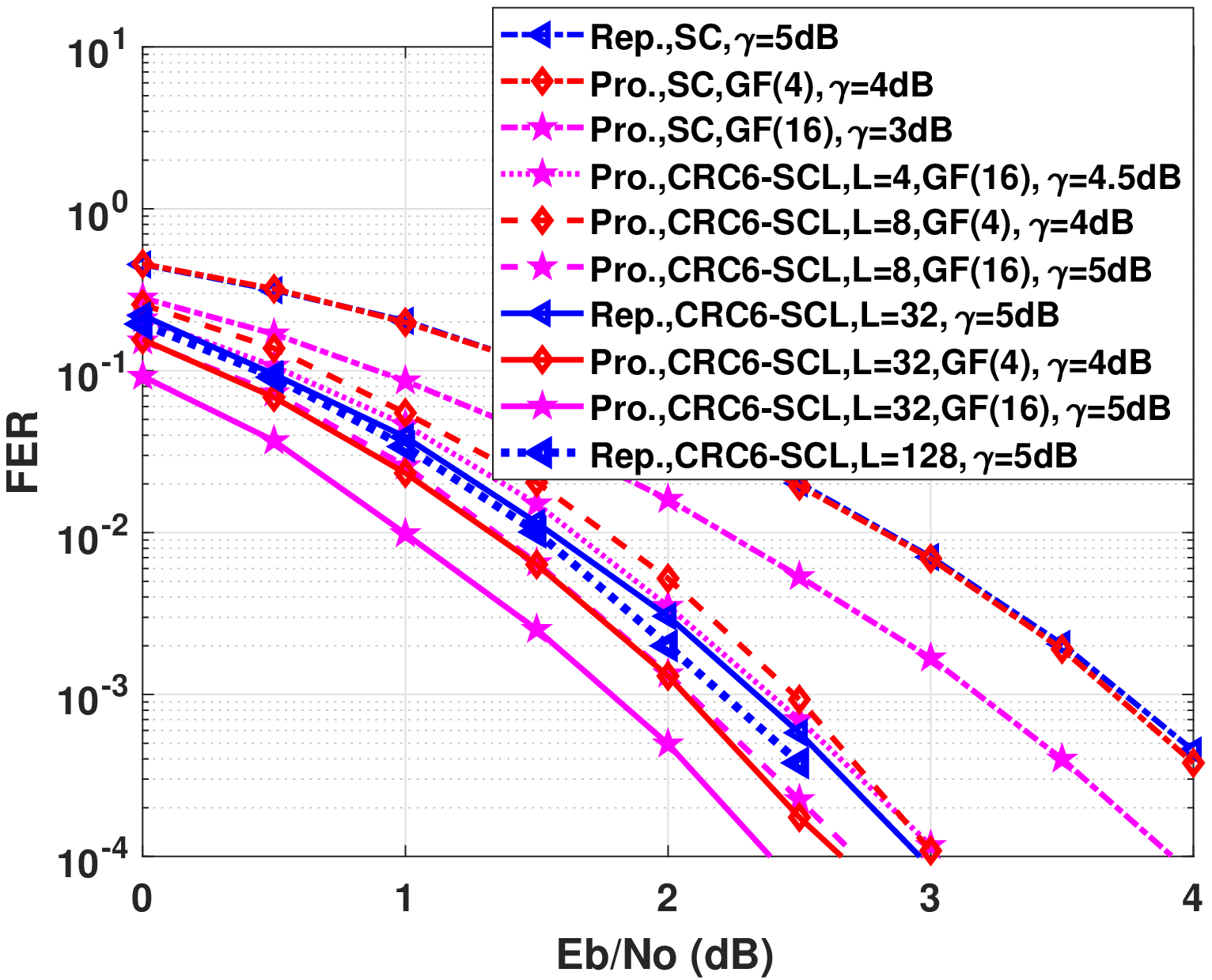}
\caption{$N=8192$, $n=256$, $k=40$, $r=32$, $t=2, 4$.}
\label{fig:256k40}
\end{subfigure}

\medskip
\begin{subfigure}[t]{0.45\textwidth}
\centering
\includegraphics[width=1.1\linewidth]{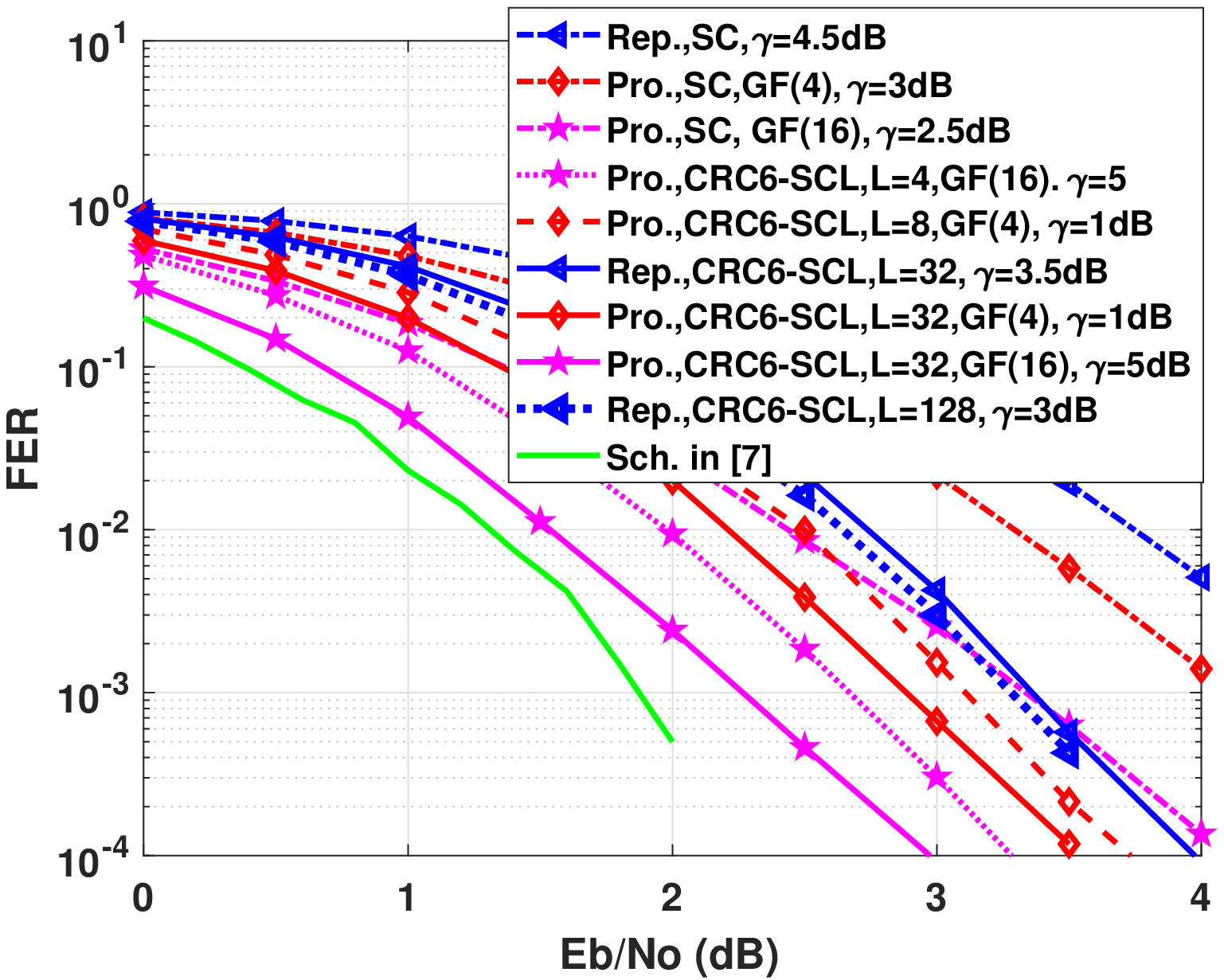}
\caption{$N=8192$, $n=128$, $k=80$, $r=64$, $t=2, 4$.}
\label{fig:128k80}
 \end{subfigure}%
\begin{subfigure}[t]{0.45\textwidth}
\centering
\includegraphics[width=1.1\linewidth]{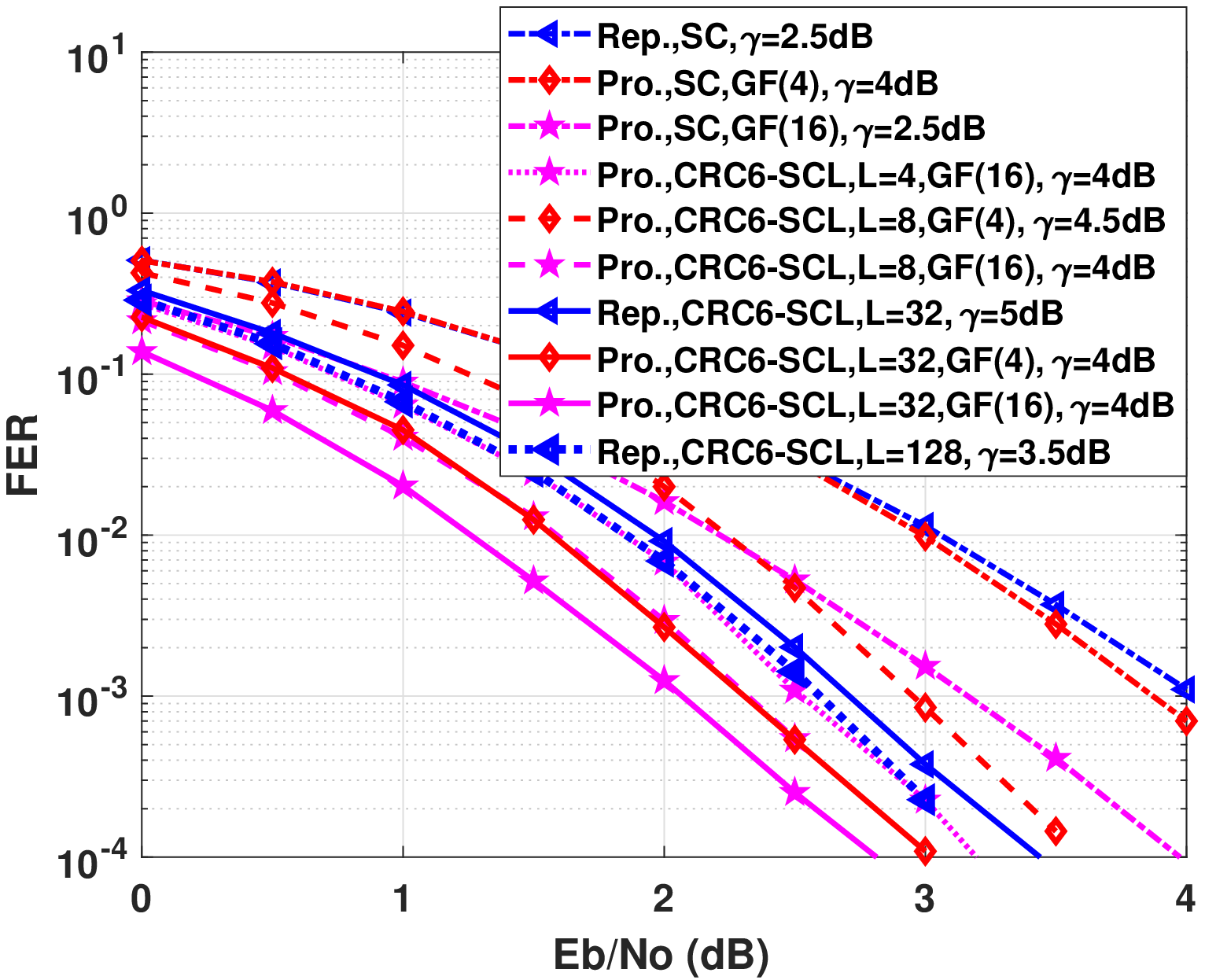}
\caption{$N=8192$, $n=128$, $k=40$, $r=64$, $t=2, 4$.}
\label{fig:128k40}
\end{subfigure}
\caption{Performance comparison of the proposed scheme with the polar-repetition scheme and the scheme in \cite{Dumerldpc} over AWGN channel for different values of $n$, $r$ and $k$. }
\end{figure*}

Figures \ref{fig:512k80}, \ref{fig:256k80}, and \ref{fig:128k80} compare the performance of the proposed scheme for $N=8192$, $k=80$, and 
over $GF(4)$ and $GF(16)$ with that of the straightforward polar-repetition scheme over AWGN channel. To illustrate the trade-off between the rate of the outer polar code and the number of the repetitions, we consider three different values $r=16, 32, 64$. The optimized design SNR $\gamma$ for each simulated case is specified in the legend of the corresponding plot.

It can be seen that the proposed scheme over $GF(16)$ outperforms the one over $GF(4)$ and the repetition scheme, under SC and CRC-aided SCL decoder with the same list size and $6$-bit CRC. For $r=16$ under CRC-aided SCL, the performance of the proposed scheme over $GF(16)$ with $L=8$ is almost the same as that of the proposed scheme over $GF(4)$  with $L=32$ and the polar-repetition scheme with $L=128$. For $r=32$ under CRC-aided SCL, the proposed scheme over $GF(16)$  with $L=8$ outperforms the proposed scheme over $GF(4)$  with $L=32$ and that one outperforms the repetition scheme  with $L=128$. Moreover, the performance of the proposed scheme over $GF(16)$ with $L=4$ is almost the same as that of the repetition scheme  with $L=128$. As we increase the number of the repetitions to $64$, due to the loss in the asymptotic achievable rate, the performance of polar-repetition scheme degrades. As a result, the performance of the proposed scheme over $GF(4)$ with $L=8$ outperforms that of the repetition scheme  with $L=128$.

For comparison, the performance of the scheme proposed in \cite{Dumerldpc} for low-SNR regime is also considered. 
It can be seen that for $r=16$, the proposed schemes over $GF(4)$ under CRC-aided SCL with $L=32$ and $GF(16)$ with $L=8$ outperform the scheme in \cite{Dumerldpc}. Moreover, the performance of the scheme in \cite{Dumerldpc} is the same as that of the proposed scheme over $GF(4)$  with $L=8$ and repetition scheme with $L=32$. For $r=32$, the proposed scheme over $GF(16)$  with $L=8$ outperforms the scheme in \cite{Dumerldpc}. On the other hand, the performance of the proposed scheme over $GF(4)$  with $L=32$ is almost the same as that of \cite{Dumerldpc}. However, for $r=64$, the scheme in \cite{Dumerldpc} outperforms the proposed scheme over $GF(16)$ with $L=32$.

Figures \ref{fig:512k40}, \ref{fig:256k40}, and \ref{fig:128k40} compare the performance of the proposed scheme over $GF(4)$ and $GF(16)$ with that of the straightforward polar-repetition scheme over AWGN channel for $N=8192$, $k=40$. Similarly, the proposed scheme over $GF(16)$ outperforms the one over $GF(4)$ and repetition scheme under both SC and CRC-aided SCL decoder. 
The scheme in \cite{Dumerldpc} can not be realized for these parameters.

\begin{figure*}[tp!]
     \centering
    \begin{subfigure}[t]{0.45\textwidth}
         \centering
         \includegraphics[width=\textwidth]{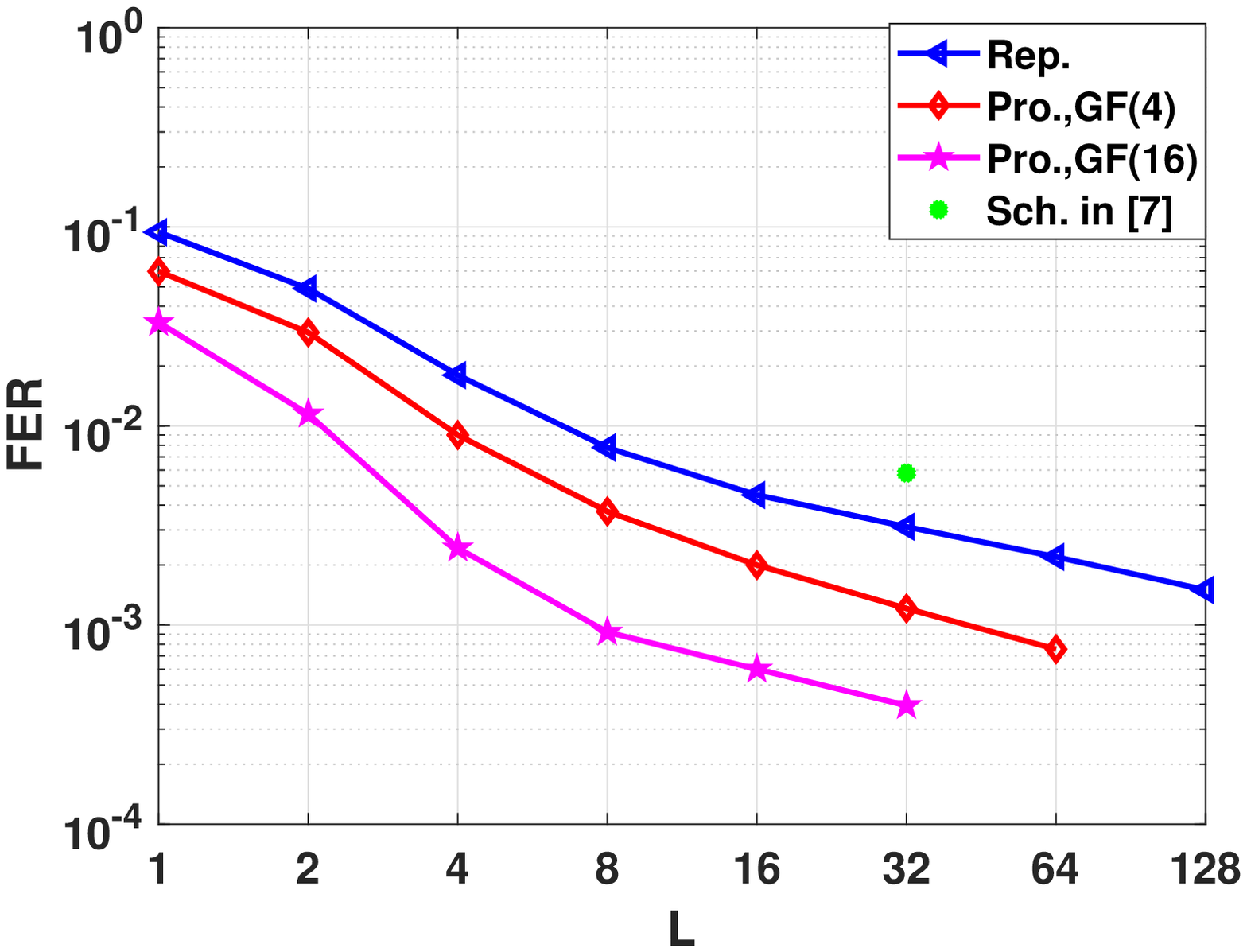}
         \caption{Performance}
         \label{Performance512}
     \end{subfigure}%
    \begin{subfigure}[t]{0.45\textwidth}
         \centering
         \includegraphics[width=\textwidth]{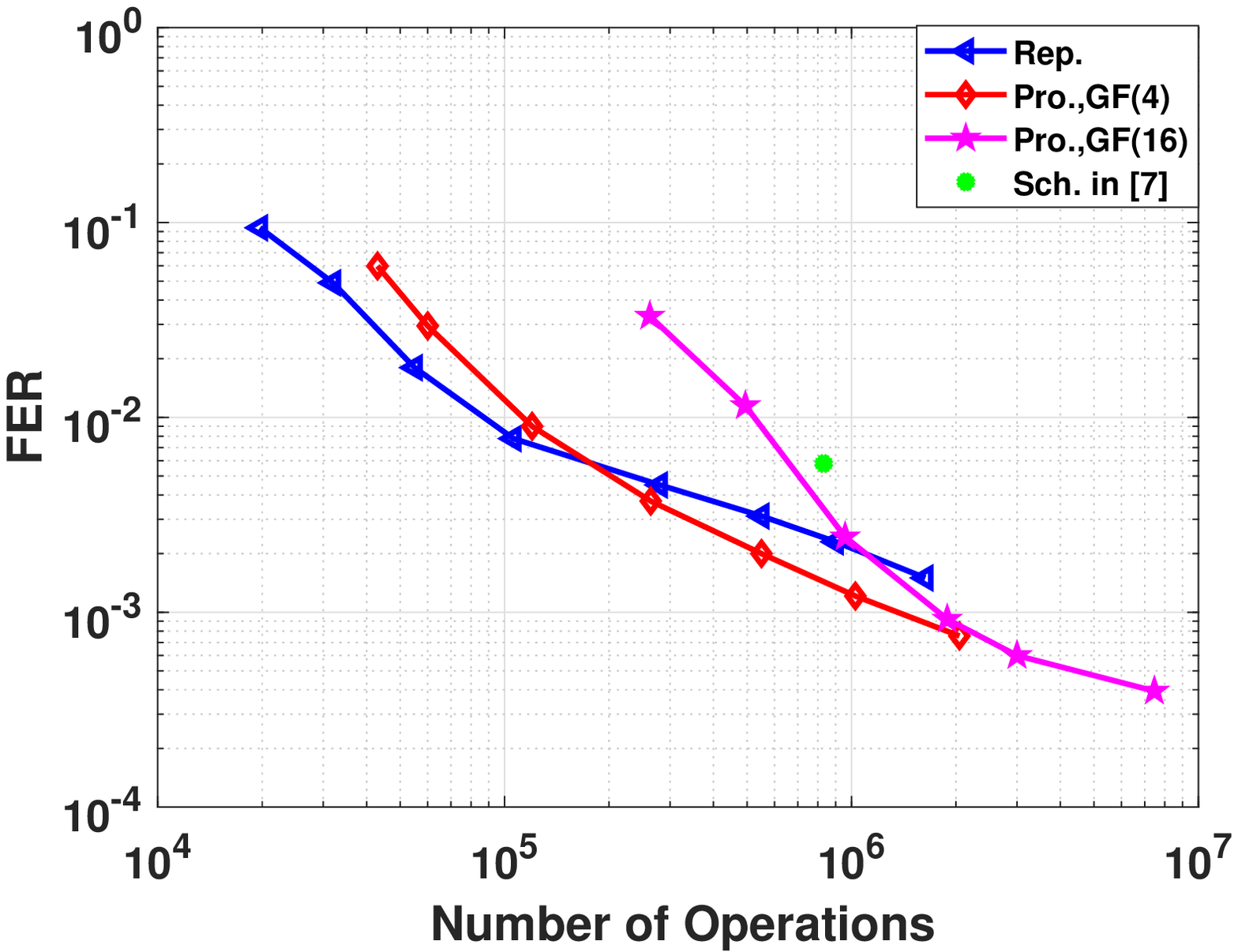}
         \caption{Complexity}
                  \label{Complexity512}
        \end{subfigure}
        \caption{CRC-aided SCL decoding over AWGN channel at $Eb/No = 1.5 $ dB for $N=8192$, $n=512$, $r=16$, $k=80$.}

        \medskip
         \begin{subfigure}[t]{0.45\textwidth}
         \centering
         \includegraphics[width=\textwidth]{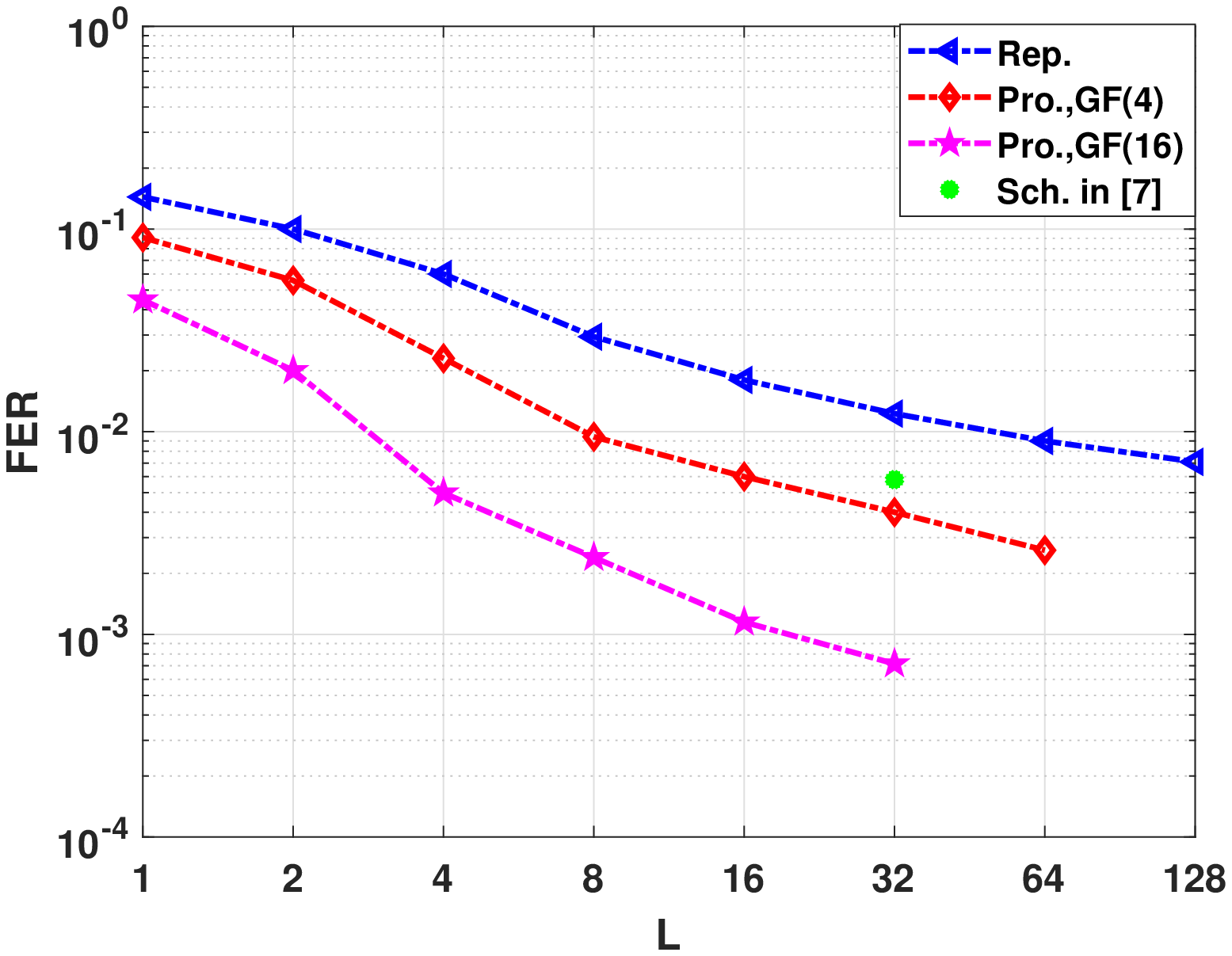}
         \caption{Performance}
                  \label{Performance256}
\end{subfigure}%
    \begin{subfigure}[t]{0.45\textwidth}
         \centering
         \includegraphics[width=\textwidth]{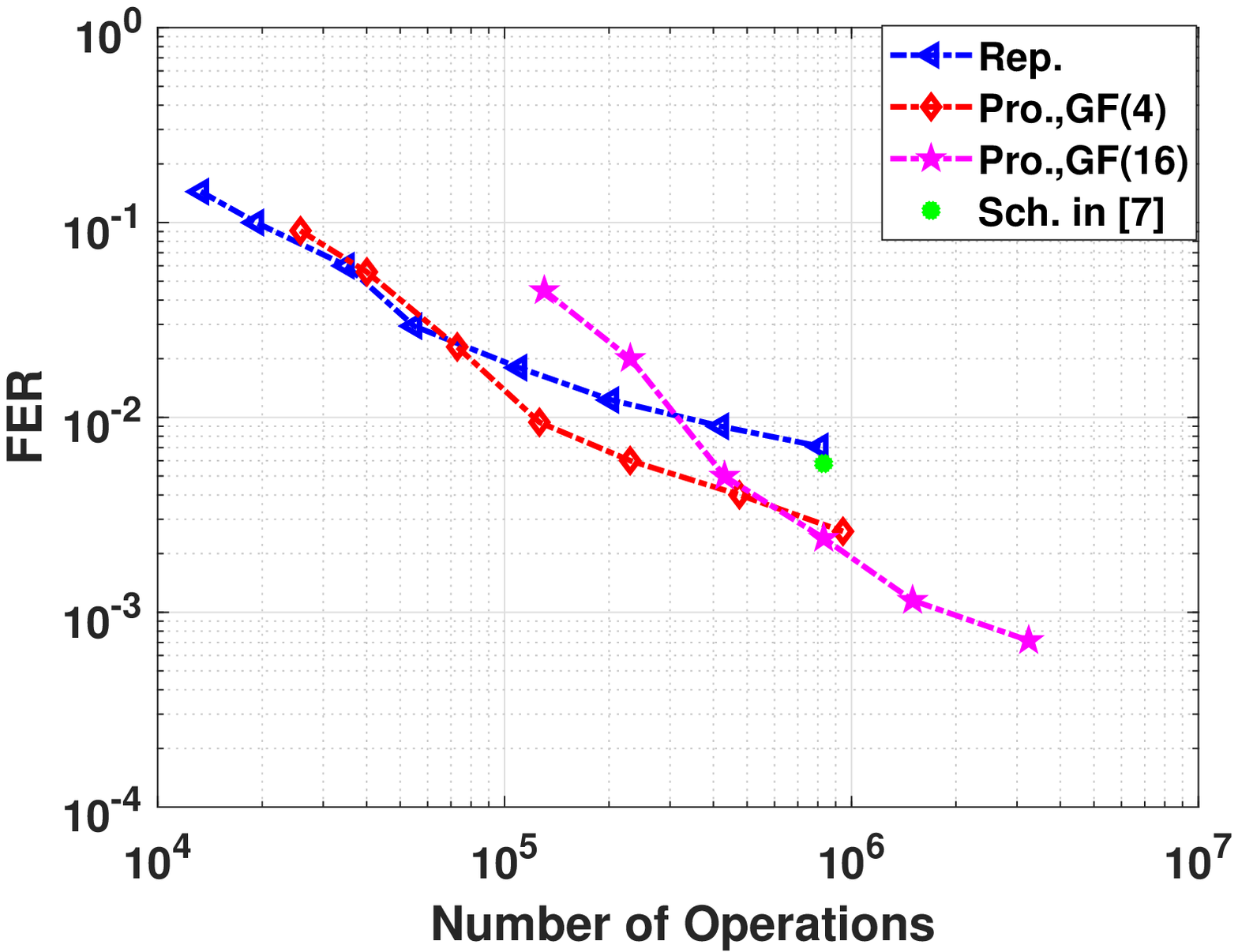}
         \caption{Complexity}
                \label{Complexity256}
        \end{subfigure}
         \caption{CRC-aided SCL decoding over AWGN channel at $Eb/No = 1.5$ dB for $N=8192$, $n=256$, $r=32$, $k=80$.}

        \medskip
         \begin{subfigure}[t]{0.45\textwidth}
         \centering
         \includegraphics[width=\textwidth]{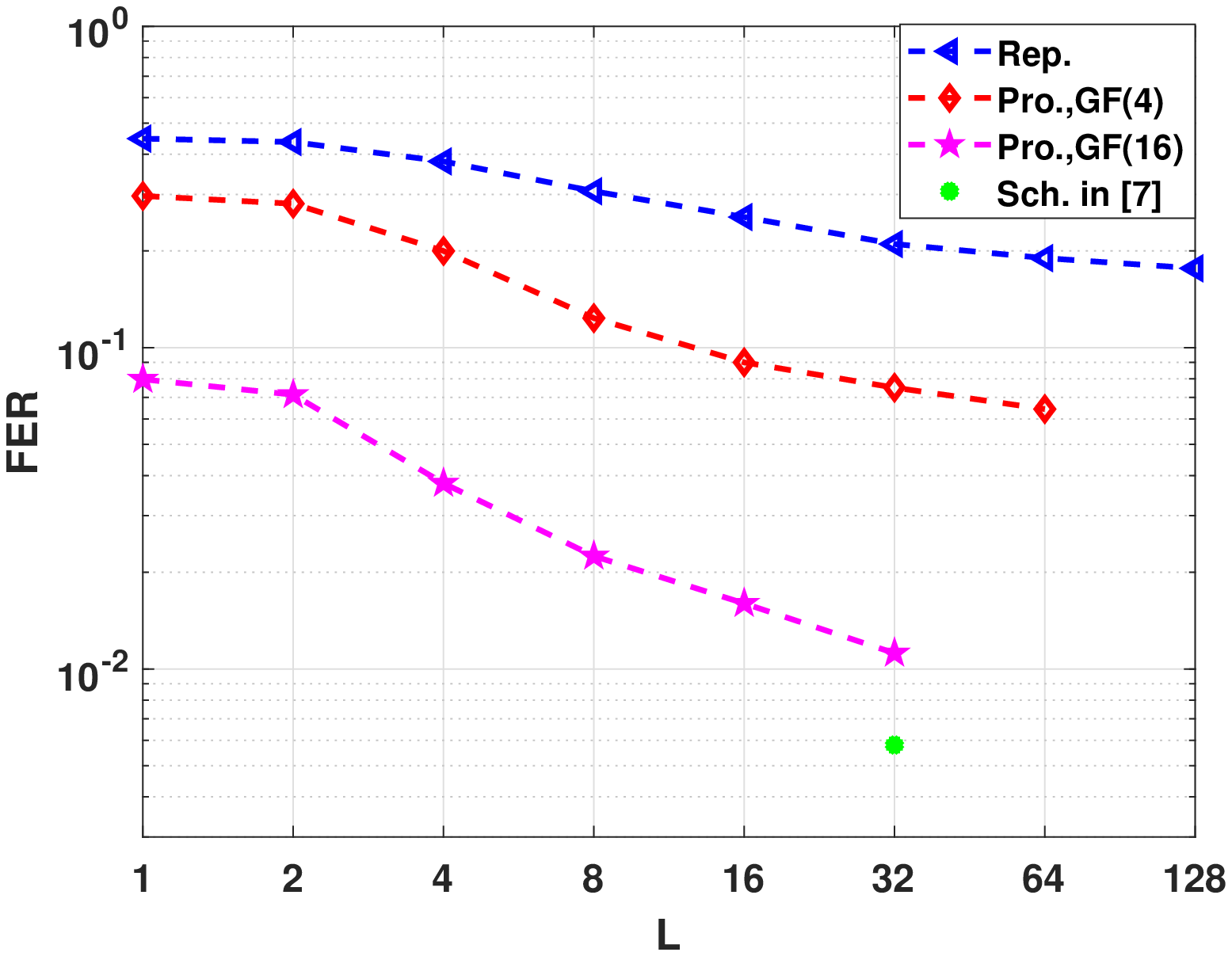}
         \caption{Performance}
         \label{Performance128}
\end{subfigure}%
    \begin{subfigure}[t]{0.45\textwidth}
         \centering
         \includegraphics[width=\textwidth]{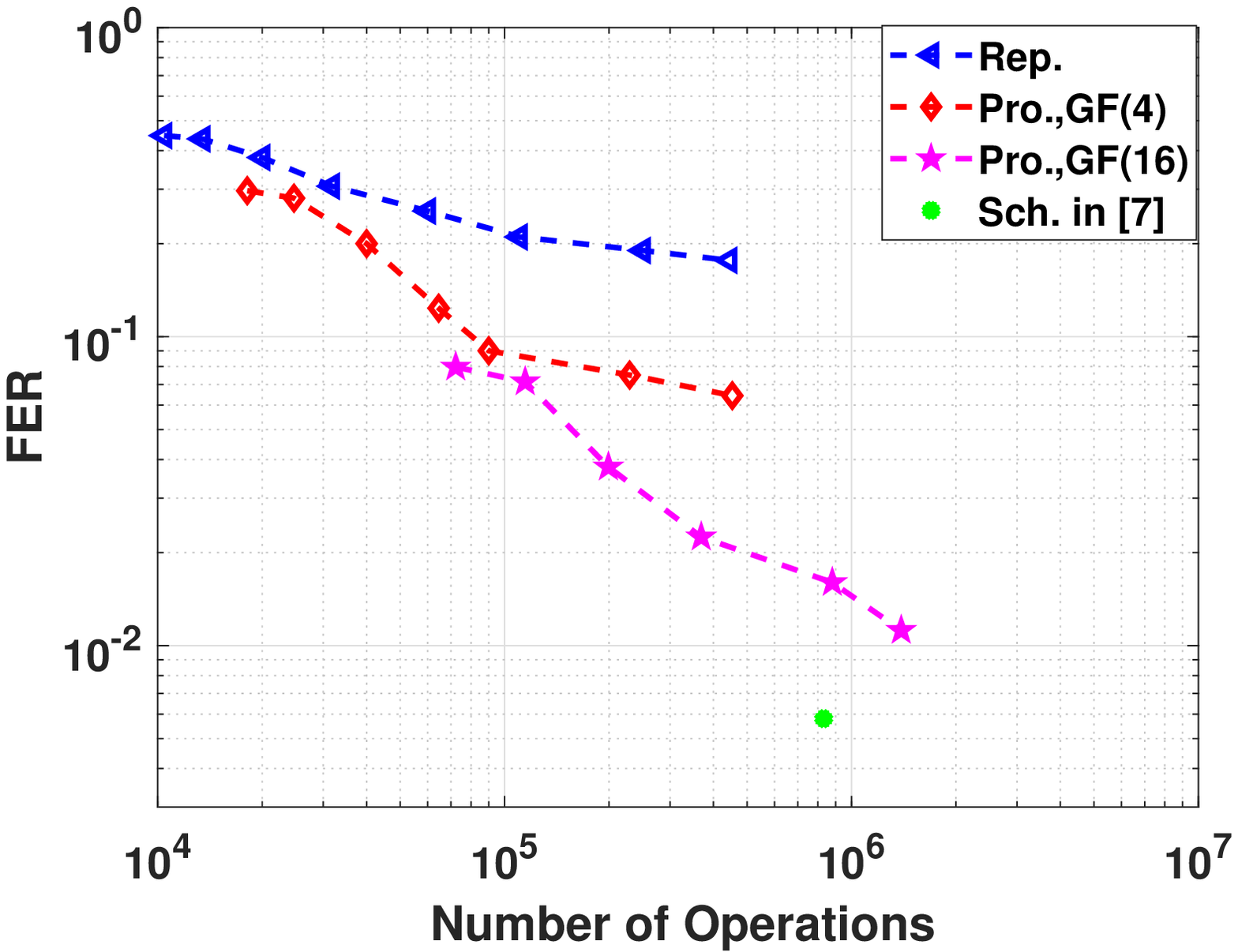}
         \caption{Complexity}
                  \label{Complexity128}
        \end{subfigure}
         \caption{CRC-aided SCL decoding over AWGN channel at $Eb/No = 1.5$ dB for $N=8192$, $n=128$, $r=64$, $k=80$.}

\end{figure*}
Figures \ref{Performance512}, \ref{Performance256} and \ref{Performance128} present the simulation results of the hybrid non-binary repeated polar code and the polar repetition scheme under CRC-aided SCL decoder as a function of  $L$ over AWGN channel at $Eb/N0 = 1.5$ dB for $N=8192$, $k=80$ and three different number of repetitions, $r=16, 32, 64$, respectively. It can be seen that the hybrid non-binary repeated polar code over $GF(16)$ requires significantly lower list size $L$ to achieve the same performance as the polar-repetition scheme over $GF(4)$. Moreover, this gap grows with $L$. For comparison, the performance of the scheme proposed in \cite{Dumerldpc} with $L=32$ is also provided.

Figure \ref{Complexity512} presents the performance of the CRC-aided SCL decoding of the schemes in Figure \ref{Performance512} in terms of the decoding complexity. Observe that the proposed scheme over $GF(4)$ can provide better performance than polar-repetition scheme with the same decoding complexity for FER $\leq 6 \times 10^{-3}$ ($L> 4$ for the proposed scheme over $GF(4)$ and $L>8$ for polar-repetition scheme). This is due to the higher slope of the corresponding curve in Figure \ref{Performance512}, which eventually enables one to compensate for the relatively high complexity of the LLR computation presented in Table \ref{complexitycomparison}. The proposed scheme over $GF(16)$ has greater processing complexity than the one over $GF(4)$, so that its curve intersects the polar-repetition scheme at FER $=2 \times 10^{-3}$ ($L\geq 4$ for the proposed scheme over $GF(16)$ and $L\geq 64$ for polar-repetition scheme). Moreover, the proposed schemes over $GF(4)$ and $GF(16)$ have the same FER $= 8 \times 10^{-4}$ for $L\geq 64$  and $L > 8$, respectively. Similar behaviour is observed for $r=32$ repetitions in Figure \ref{Complexity256}. However, as we increase $r$ from $16$ to $32$, the decoding complexity decreases, while the performance of all the schemes degrades. As a result, the intersections of the curves happen at lower FER and smaller number of the operations. For $r=64$ repetitions in Figure \ref{Complexity128}, a significant degradation in the performance occurs due to the reduced achievable rate of the outer polar code due to having a shorter length. For any $L$, the hybrid non-binary repeated polar code over $GF(16)$ provides the best performance for the same decoding complexity.

It can be observed that for $r=16$ repetitions (see Figure \ref{Complexity512}), the hybrid non-binary repeated polar code and polar-repetition scheme outperform the scheme in \cite{Dumerldpc} with the same decoding complexity. For $r=32$  repetitions (see Figure \ref{Complexity256}), the scheme in \cite{Dumerldpc} slightly outperforms the polar-repetition, but not the proposed scheme for the same decoding complexity. Only for $64$ repetitions (see Figure \ref{Complexity128}), the scheme in \cite{Dumerldpc} outperforms the others with the same decoding complexity. 

\begin{figure}[h]
\centering
\includegraphics[width=1.1\linewidth]{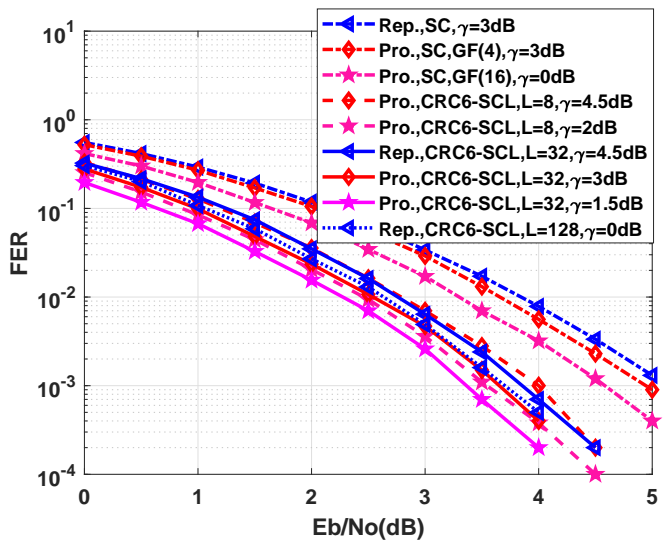}
\caption{Performance comparison of the proposed scheme with the polar-repetition over Rayleigh fading channel for $N=8192$, $n=512$, $k=80$, $r=16$, $t=2, 4$.}

\label{fig:fading}
\end{figure}

Figure \ref{fig:fading} compares the performance of the proposed scheme for $N=8192$, $k=80$, $r=16$ over $GF(4)$ and $GF(16)$ with that of the straightforward polar-repetition scheme over Rayleigh fading channel. For each repetition block, we consider one distinct fading coefficient. As a result, in each frame of size $8192$, there are $16$ distinct fading coefficients and each block of length $512$ symbols in each repetition will be affected by the same fading coefficient.
Similar to AWGN channel, it can be seen that the proposed scheme over $GF(16)$ outperforms the one over $GF(4)$ and the repetition scheme, under SC and CRC-aided SCL decoder with the same list size and $6$-bit CRC. 

\subsection{Performance Analysis}

In this subsection, we analyze the performance of the proposed scheme to gain insight into its better performance under the CRC-aided SCL decoder. 
Since the SCL decoder with large list size achieves performance very close to that of the  ML decoder, we study the performance of the proposed scheme under ML decoding to have better analytical understanding.

The block error probability under ML decoding can be estimated via the truncated union bound as follows, \cite{Sason}.
\begin{equation}
P_e^{\textup{ML}} \leq \sum_{i=d}^{n} A_i Q(\sqrt{2 i R E_b/N_o)},
\end{equation}
where $A_i$ is the number of the codewords of weight $i$ and $d$ is the minimum distance of the code.
At high SNR, upper bound on $P_e^{\textup{ML}}$ depends primarily on $d$ and $A_d$. Hence, to obtain a good performance under ML decoding, one needs to eliminate low-weight non-zero (LWNZ) codewords from the code. 

To enumerate the LWNZ codewords, we transmit the all-zero codeword in the extremely high SNR regime under SCL decoder with very large list size, \cite{Li}. In this case, it is expected that the list most likely contains only the codewords with the least Hamming weights. 

Table \ref{lowweightcodewords} compares the number of low-weight codewords of the proposed scheme with the polar-repetition ones over AWGN channel for $N=8192$, $n=512$, $r=16$, $k=40$.
As it is expected the number of LWNZ codewords for the proposed scheme over $GF(4)$ is less than that of the polar-repetition scheme. For $GF(16)$, although the minimum distance is slightly reduced, the number of low-weight codewords is still much smaller compared to the polar-repetition scheme.

\begin{table*}
\begin{center}
\scalebox{0.7}{
\begin{tabular}{|P{3cm}|P{3cm}|P{2.2cm}|P{1.5cm}|P{2.2cm}|P{1.5cm}|P{2.2cm}|P{1.5cm}|P{2.2cm}|P{1.5cm}|}
 \hline
 \multirow{1}{*}{Parameters} & Schemes &$A_{1000}- A_{1024}$ &$A_{64\times 16}$ & $A_{1329}-A_{2047}$& $A_{128 \times 16}$ & $A_{2049}-A_{3071}$ & $A_{192 \times 16}$ & $A_{3073}-A_{4095}$&$A_{256 \times 16}$\\
\hline
   \multirow{7}{*}{$n=512$, $r=16$ } & Prop. Sch., $t=2$ & $0$ & $0$ & $<10$ & $0$ & $<30$ & $3$ & $<230$ & $230$\\
   \cline{2-10}
   \multirow{7}{*}{$k=40$} & Prop. Sch., $t=4$ & $2$ & $0$ &$32766$& $0$ &$0$& $0$ & $0$ & $0$\\
\cline{2-10}
    & Rep. Sch. &$0$ & $105$ &$0$& $1365$ &$0$& $5005$ & $0$ & $22819$\\ 
\cline{2-10}
&  & $A_{4097}-A_{5119}$& $A_{320\times 16}$ & $A_{5121}-A_{6143}$& $A_{384 \times 16}$ & $A_{6145}-A_{7167}$ & $A_{448 \times 16}$ & $>A_{7169}$ &\\
\cline{2-10}
 & Prop. Sch., $t=2$ &  $<230$ & $0$& $<20$ & $0$ & $0$ & $0$ & $0$& \\
 \cline{2-10}
 & Prop. Sch., $t=4$ &  $0$ & $0$& $0$ & $0$ & $0$ & $0$ & $0$ &\\
\cline{2-10} 
 & Rep. Sch. & $0$ & $3003$ &$0$& $455$ &$0$& $15$ & $0$ &\\ 
\hline
\end{tabular}%
}
\end{center}
\caption{Number of Low-Weight Codewords}
\label{lowweightcodewords}
\end{table*}



\section{Conclusion}
In this paper, we proposed a new concatenation scheme to improve upon the performance of the polar-repetition scheme in the low-SNR regime. The proposed scheme is the concatenation of a hybrid non-binary polar code with a multiplicative repetition code. Extensive simulation and numerical results show that the proposed scheme provides a better trade-off between the decoding complexity and performance compared with the polar-repetition scheme, under CRC-aided SCL decoder over AWGN and Rayleigh fading channels.

There are several directions for the future research. Finding the best permutation for the inner non-binary multiplicative repetition code is an interesting problem. Reducing the decoding complexity of our proposed constructions by pruning the decoding trellis and simplifying it \cite{Alamdaryazdi}, and reducing the latency to become sub-linear \cite{MondelliLatency} are other interesting directions. Also, analyzing the polarization properties of the proposed scheme with the optimal permutation in terms of error exponent, \cite{AT}, and scaling exponent, \cite{HassaniSE}, \cite{MITI2020}, and comparing them with polarization based on kernels constructed in \cite{Lin}--\cite{FaribaTVT2020}, are other interesting open problems.

\section*{Acknowledgment}
The authors would like to thank Navid Gharavi and Prof. Ilya Dumer for providing the simulation results of their proposed scheme in \cite{Dumerldpc}.


\end{document}